\documentclass[a4paper]{article}

\usepackage{amssymb}

\usepackage[figuresright]{rotating}

\DeclareRobustCommand{\rchi}{{\mathpalette\irchi\relax}}
\newcommand{\irchi}[2]{\raisebox{\depth}{$#1\chi$}} 
\newcommand{\kcep}{$\rchi$-CEP\xspace}

\usepackage{hyperref}

\usepackage{xcolor}
\usepackage{xspace}
\usepackage[normalem]{ulem}

\newcommand{\scep}{\emph{SCEPter}\xspace}

\usepackage{hyperref}
\usepackage{graphicx} 
\graphicspath{{./}}
\DeclareGraphicsExtensions{.jpg,.pdf,.png}

\usepackage{float}
\usepackage{caption}
\DeclareCaptionType{copyrightbox}
\usepackage{subfig}

\usepackage{algorithm}
\usepackage{algorithmic}

\usepackage{anyfontsize}

\usepackage{MnSymbol}

\newtheorem{query}{\textbf{Query}}
\newtheorem{rerule}{\textbf{Rule}}

\usepackage{times}		
\newcommand\Mark[1]{\textsuperscript#1}
\newcommand{\HRule}{\rule{\linewidth}{0.5mm}}

\begin{document}

\newpage

\vspace*{0.1cm}

\begin{center}



\HRule \\[0.4cm]
{ \huge \bfseries Knowledge-infused and Consistent Complex Event Processing over Real-time and Persistent Streams}\\[0.4cm]

\HRule \\[1.5cm]

\centering
\begingroup

{
\Large Qunzhi Zhou\Mark{1}, Yogesh Simmhan\Mark{2} and Viktor Prasanna\Mark{1}\Mark{,}\Mark{3}}\\[1em]
\Large\Mark{1}Department of Computer Science, \\University of Southern California, USA\vspace{0.15cm}
\\\Large\Mark{2}Department of Computational and Data Sciences, \\Indian Institute of Science, Bangalore, India\vspace{0.15cm}
\\\Large\Mark{3}Ming Hsieh Department of Electrical Engineering, \\University of Southern California, USA\vspace{0.15cm}

\endgroup


\vspace{2cm}

\textsc{\Large Accepted in Future Generation Computer Systems, October 27, 2016}\\[0.3cm]

\end{center}

%
%
%
%
%
%
%

\newpage

\begin{abstract}
Emerging applications in Internet of Things (IoT) and Cyber-Physical Systems (CPS) present novel challenges to Big Data platforms for performing online analytics. Ubiquitous sensors from IoT deployments are able to generate data streams at high \emph{velocity}, that include information from a \emph{variety} of domains, and accumulate to large \emph{volumes} on disk. Complex Event Processing (CEP) is recognized as an important real-time computing paradigm for analyzing continuous data streams.  However, existing work on CEP is largely limited to relational query processing, exposing two distinctive gaps for query specification and execution: (1) infusing the relational query model with higher level \emph{knowledge semantics}, and (2) seamless \emph{query evaluation across temporal spaces} that span past, present and future events. These allow accessible analytics over data streams having properties from different disciplines, and help span the velocity (real-time) and volume (persistent) dimensions. In this article, we introduce a Knowledge-infused CEP (\kcep) framework that provides \emph{domain-aware knowledge query constructs} along with temporal operators that allow end-to-end queries to span across \emph{real-time and persistent streams}. We translate this query model to efficient query execution over online and offline data streams, proposing several optimizations to mitigate the overheads introduced by evaluating semantic predicates and in accessing high-volume historic data streams. In particular, we also address temporal consistency issues that arise during fault recovery of query plans that span the boundary between real-time and persistent streams. The proposed \kcep query model and execution approaches are implemented in our prototype semantic CEP engine, \scep. We validate our query model using domain-aware CEP queries from a real-world Smart Power Grid application, and experimentally analyze the benefits of our optimizations for executing these queries, using event streams from a \emph{campus-microgrid IoT deployment}. Our results show that we are able to sustain a processing throughput of $3,000$~events/secs for \kcep queries, a $30\times$ improvement over the baseline and sufficient to support a Smart Township, and can resume consistent processing within $20$~secs after stream outages as long as 2~hours.
\end{abstract}







\section{Introduction}
\label{sec:intro}

There is a growing prevalence of Internet of Things (IoT) sensors and actuators, both for specific domains such as Smart Grids and Smart Transportation, and through lifestyle devices such as Smart Watches and Fitness Bands. These sensors generate streams of events that arrive continuously, and can include observations from multiple domains that need to be analyzed. Complex Event Processing (CEP) is a computing paradigm for online analytics over such high velocity data streams~\cite{Demers2007}. Contemporary CEP systems offer the capability to specify event patterns to detect value thresholds or correlation constraints, and execute them continuously over events streams. In particular, CEP addresses the \emph{velocity} dimension of the 3-Vs of Big Data~\cite{big-data-3v}, \emph{volume} and \emph{variety} being the other two, and has grown popular for operational intelligence where online pattern detection drives real-time response. It has been used in domains varying from mobile computing~\cite{xiao:tc:2012} and financial services~\cite{Asaf2006} to healthcare~\cite{wang:vldb:2011} and sports analytics~\cite{debs-challenge-soccer}.

One emerging domain 
where CEP can prove vital is in \emph{Cyber Physical Systems} (CPS)~\cite{Poovendran2012}, which is a special case of IoT. In CPS, the operation and optimization of \emph{physical} infrastructure is based on analytics performed on \emph{cyber}-infrastructure, and typically happens in a closed-loop cycle. CPS encompasses many aspects of smart cities where diverse events on infrastructure conditions, be they about a transportation network~\cite{Biem2010} or power grid~\cite{ferc2008}, are integrated with prior knowledge of the infrastructure to offer insight on the system behavior. CEP engines can help analyze such event streams and detect event patterns that need an operational response. For e.g., detecting a traffic overflow based on events from sensors on an upstream road can cause downstream traffic signals to change, or stress on a neighborhood transformer detected by analyzing residential smart meter readings can trigger a notification requesting consumers in that community to curtail their power consumption. 

While CEP offers a useful paradigm to perform real-time analytics over event streams, there are two distinctive capabilities lacking in traditional CEP engines that are necessary for their effective use by emerging IoT applications:
\begin{enumerate}
\item IoT domains need to perform analytics over multi-disciplinary data sources for effective decision making. A CEP query model has to be expressive enough to capture such information richness while also being simple enough for end users to specify such analytics by hiding domain complexities. Traditional CEP systems require \emph{syntactic} queries to be specified over explicit properties present in the event contents, such as sensor IDs and equipment numbers. This makes CEP queries difficult to specify and manage for many numbers of diverse devices and sensors that are constantly in flux. \emph{Semantic} concepts defined by domain ontologies~\cite{Crapo, YogeshGreenIT2011} can help raise the abstraction by referring to events using concepts rather than just content. This requires the CEP query definitions to be infused with knowledge models and \emph{semantic predicates}, and further translated into \emph{efficient execution}. 

\item While CEP systems allow queries to be specified over current and future events that arrive on an event stream, event analytics may require the correlation between events that happen in the present as well as in the past. Traditional CEP systems do not allow queries to be specified \emph{after} an event has occurred to match the past event, necessary for exploratory analytics. Secondly, even after a query is specified, CEP systems can be memory-intensive when processing queries with a large time window, some as wide as days. This motivates the need for \emph{lazy definition} of queries after an event has happened, and their \emph{consistent and scalable execution} over end-to-end event streams that span past events persisted to disk and real-time events that arrive over the network. 
\end{enumerate} 

In other words, while current CEP systems support analytics over high \emph{velocity} event data, we also need to support data \emph{variety} in the form of diverse domains concepts present in IoT event streams, and analysis over large \emph{volumes} of archived event streams, thus uniquely encompassing all three dimensions of Big Data.

Most existing CEP systems expose gaps in their ability to model queries at a higher-abstraction, and their execution on end-to-end event streams. These systems process relational events with syntactic queries directly defined on their properties, and expose users to the underlying events' structural heterogeneity~\cite{Demers2007}. Recently, C-SPARQL \cite{Barbieri2010} and ETALIS \cite{Anicic2012} introduced event context or semantics into CEP for abstract query specification, allowing background knowledge to be combined with real-time events. But these are solutions that rely on Semantic Web technologies, leveraging inference engines to model and query over events and domain knowledge. As a result, common CEP temporal patterns such as klenee closure and matching policies for event selection and consumption~\cite{Cugola2012Survey, Etzion2010} are not supported. Further they lack the scalability of CEP systems that are optimized for pattern matching over streaming events. On the other hand, existing CEP systems focus on processing real-time events, without considering archived event streams. Integrated querying over real-time and persistent data has attracted some interest from active databases~\cite{Adaikkalavan2006}. These leverage triggers in relational query engines to process time varying data that is persisted. DataCell~\cite{Liarou2009} layers in-memory tables on top of database kernels to handle online queries. These \emph{database-centric} systems sacrifice the expressivity of CEP temporal query patterns and introduce additional latency into matched results. A recency-based CEP model~\cite{Dindar2011} supports a \emph{happened-before} relation which links live streams with persisted events. However, this is limited to correlating patterns between real-time and archived streams, rather than seamlessly detecting patterns that span across both. Big Data platforms that perform in-memory computing, like Apache Spark Streaming, support high-throughput incremental processing over persisted data by loading batches of datasets into memory. But Spark treats the data content as opaque and users need to implement all processing logic over them. Further, while such systems are faster than batch processing platforms like Hadoop, they still have a higher latency than CEP engines.

In this article, we propose \scep, a Knowledge-infused CEP (\kcep, pronounced \emph{kai}-CEP) framework which uniformly processes queries across persistent and real-time event streams, end-to-end, and addresses the gaps identified above. \scep is motivated by and evaluated within the Smart Power Grid CPS domain, as part of the US Department of Energy sponsored Los Angeles Smart Grid project~\cite{simmhan:cise:2013}. \scep allows users to specify expressive event patterns using semantic concepts over heterogeneous information sources, permits lazy-specification of queries for online and \emph{post facto} analytics, ensures temporally consistent execution across end-to-end event streams, and includes optimizations to mitigate performance overheads introduced by such features. 

Specifically, our contributions are as follows:
\begin{enumerate} 
\item We take a CEP-centric approach to infuse knowledge-models and domain semantics into relational CEP events (\S~\ref{sec:eventpattern}). Further, we propose a unified \kcep query model that supports \emph{semantic predicates} that leverage both real-time event data and static knowledge-bases, and \emph{temporal predicates} that can operate end-to-end over real-time and persistent event streams.
\item We discuss query processing techniques for real-time event streams (\S~\ref{sec:streamscep}), persistent event archives (\S~\ref{sec:dbprocessing}), and executions that span the temporal boundary between the two (\S~\ref{sec:integrated}). We propose \emph{performance optimizations} like event buffering and semantic caching for real-time querying, and hybrid rewriting and replay for archive processing to offer low latency, consistency and resiliency in the presence of temporal gaps in event streams.
\item We implement the proposed \kcep query model and execution techniques within \scep (\S~\ref{sec:arch}). Further, we validate its effectiveness in representing event analytics from the Smart Power Grid CPS domain (\S~\ref{sec:background}), and empirically evaluate the performance of \scep for different input event rates, buffer and cache capacities, and temporal stream gaps, using real event data and queries from Smart Grid applications (\S~\ref{sec:experiments}).
\end{enumerate}

\section{Background, Problem Motivation and Approach}
\label{sec:background}
With the growing pressures of global warming and the critical role that electricity plays in our society, the importance of efficient and reliable operation of power grids cannot be overstated. Smart Power Grids are an exemplar of Cyber Physical Systems (CPS) and Internet of Things (IoT), and form a key building-block for the push toward smart and sustainable cities. Smart Grids integrate sensors, actuation and communication devices at the generation, transmission and distribution networks of the power grid, and utilize operational analytics over real-time information from these sensors along with static knowledge to drive intelligent management of the grid~\cite{Ramchurn:2012:cacm}.

\emph{Demand response optimization (DR)} is a key Smart Grid application that attempts to prevent a mismatch between the generation capacity and the electricity load from consumers that can cause brown-outs and black-outs. DR predicts and detects such demand-supply gaps, and curtails energy usage through shifting, shaving and shaping strategies enacted though direct control of equipment and through notifications and incentives offered to consumers. In addition to improving grid reliability, such control strategies and behavioral suggestions also ensure that we save on the cost of building additional generation capacity to meet occasional peak demand, thereby reducing the carbon footprint. 

Traditionally, DR mismatch predictions are made days ahead, statically, using seasonal averaging models over historical consumption data from across all customers in the utility area. Curtailment strategies similarly use static means such as using time-of-use pricing to encourage energy curtailment during historically high-demand periods~\cite{ferc2008}. But the ability to collect real-time power consumption data from smart meters at the consumer is allowing for \emph{dynamic DR decisions}, where predictions are done using real-time energy consumption data and curtailment strategies target individual customers with specific usage profiles~\cite{aman:2015:smartgridcomm}. However, even such time-series forecasting models need to be adjusted for outlier events that may occur, and curtailment strategies need to be responsive to real-time opportunities.

CEP is a natural tool to analyze events generated from diverse sensors, that go beyond just smart meters, and facilitate intelligent dynamic DR decisions. Building Area Networks (BAN) are IoT deployments that are able monitor the operation of electrical equipment. They observe properties like air speed, ambient light levels, and passive infra-red monitoring of human presence for infrastructure like Heating, Ventilation and Air Conditioning (HVAC) units, lighting and elevators, and physical rooms. The University of Southern California's (USC) campus microgrid IoT testbed~\cite{simmhan:cise:2013}, part of the US Department of Energy-funded Los Angeles Smart Grid Project, is one such example. At USC, over $50,000$ sensors monitor equipment spread across $115$ buildings, sampled every few seconds to a few minutes, and stream observation over the campus local area network.  CEP can help detect specific patterns of events over thousands of such event streams that may predict a power consumption situation on campus, or offer a power curtailment opportunity. 

However, as the types of situations that need to be recognized get more sophisticated and cross-domain information becomes available, such as organizational schedules and weather data, syntactic CEP engines are inadequate. As mentioned before, two of these gaps are, \textbf{(1)} Incorporating domain knowledge into CEP patterns, and \textbf{(2)} Supporting seamless CEP pattern detection over real-time and historical events. These gaps exist both in modeling and in efficient execution -- models that cannot translate to efficient frameworks are impractical while \emph{ad hoc} implementations without formal models are not generalizable. We motivate these needs using illustrative examples from Dynamic DR applications within the USC campus microgrid.

Firstly, the microgrid includes diverse infrastructure such as sensors and online services that continuously emit time-series events on equipment (e.g. \textit{AmbientTemperature} and \textit{AirflowReport} events), power consumption (\textit{MeterUpdate} event), local weather (e.g. \textit{HeatWave} event) and consumer activities (e.g. \textit{ClassSchedule} and \textit{RoomOccupancy} events). Events from the same source may vary in their format and meaning, and different sources may generate the same type of event but with different formats. New static knowledge-bases (e.g., user surveys, organizational charts) and streaming sources (e.g., mobile app feedbacks, environmental monitoring sensors) will come online as the stakeholders expand. Given the multi-disciplinary users and the need to synthesize intelligent actions, a mere structural mapping of event formats to a normalized schema is inadequate. Relating real-time events with existing knowledge-base concepts is necessary for users, such as utility managers, facility coordinators and end-use consumers, to easily define meaningful patterns~\cite{Crapo, YogeshGreenIT2011}. Infusing such knowledge awareness into pattern specification will offer a higher-level abstraction, for e.g., by specifying a match for \texttt{KWhEvent}s from \texttt{MeetingRoom}s, rather than events from sensor IDs \texttt{76284} or \texttt{35143}. This makes it faster to adapt with evolving data sources and concepts, and while ensuring scalability at runtime. 

Secondly, given the exploratory needs of this emerging domain, not all query patterns may have been defined \emph{a priori} before events of interest occur. Lazy-definition of query patterns will be common as analysts try out ``what if'' scenarios that require event queries to be applied back in time. For e.g., we may wish to specify a query that matches a power spike event at $10$AM today with similar spikes at $10$AM in the last 4 days to determine if a workday pattern is present. Further, the operational needs of DR may not tolerate failures in the CEP system that miss patterns due to intermittent hardware or software faults at runtime. This, combined with the fact that events in a Smart Grid are often archived for regulatory compliance~\cite{martinez:pnnl:2005} and data mining~\cite{aman2011DDDM}, means that there is an opportunity to enhance the robustness and flexibility of the CEP system if it can seamlessly perform the same query across both real-time and persisted events. This requires translating a temporal pattern definition into queries that operate both exclusively on real-time event streams and can seamlessly span in-memory and on-disk event streams, while executing them with low latency.

Consider a DR scenario from the USC microgrid. 

\emph{A facility manager in the microgrid needs to reduce the electricity load on campus by $10$\% in response to a request received at \textbf{10AM} from the power utility asking to curtail the consumption from noon onwards. In response, the manager wants to evaluate the potential reduction in load by reducing the fan speed of HVAC units of \textbf{Lecture Halls} where the \textbf{airflow} of the unit exceeds \textbf{500~cfm}. However, she wants this pattern to be \textbf{detected from 9AM}.} 

Here, \emph{Lecture Hall} is a semantic concept not present in the raw sensor data but available in a microgrid knowledge-base that links the HVAC unit to its room location; \emph{airflow} is a normalized concept that describes event attributes that may be syntactically named `flowrate' or `airvolume'; and the historical application of this pattern (from 9AM to the present, 10AM, and continuing till noon) helps to identify and curtail rooms that have been over-cooled for more than an hour.

\begin{figure}[t]
  \centering
  \includegraphics[width=2.6in]{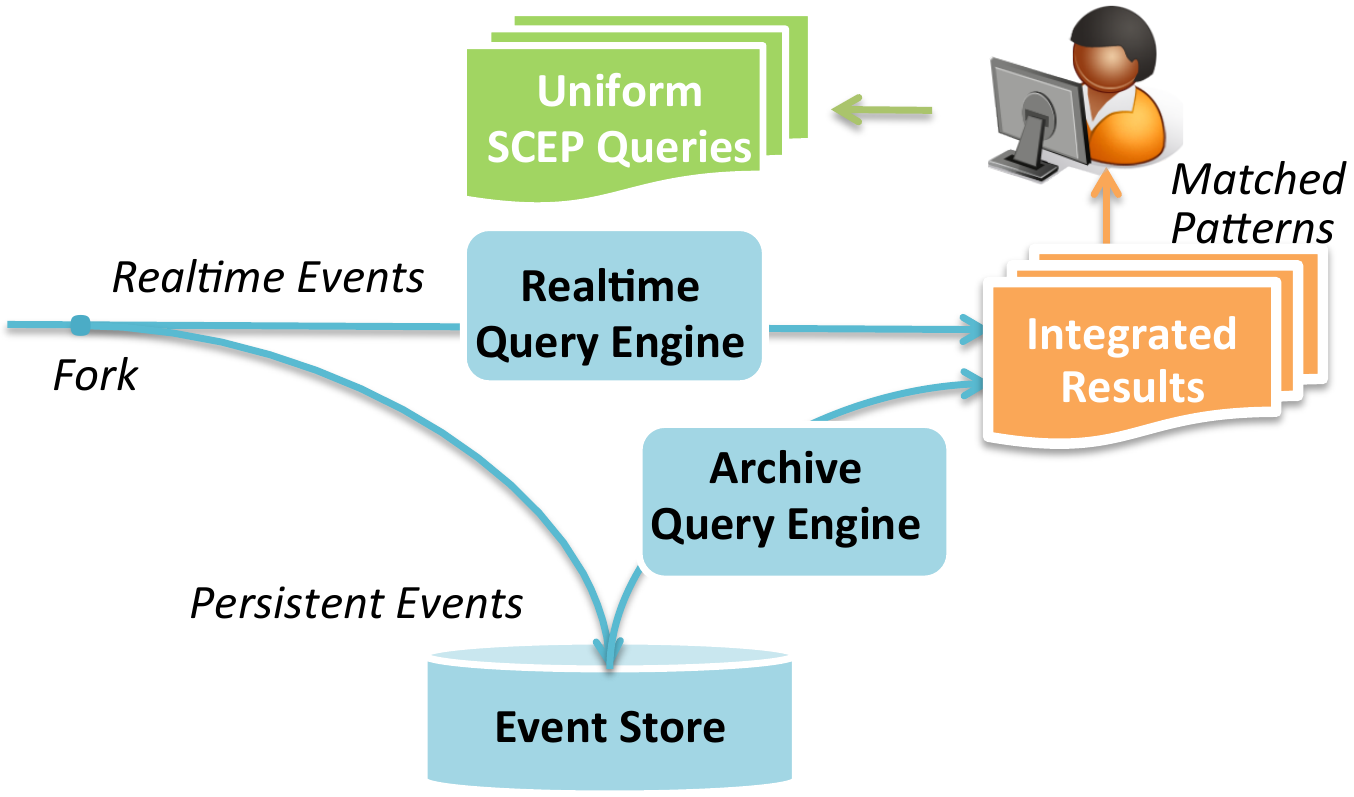}
  \caption{\kcep Querying over End-to-End Event Streams}
  \label{fig:dataflow}
\end{figure}

Fig. \ref{fig:dataflow} depicts the general scenario of integrated \kcep querying across end-to-end real-time and archived event streams. Events arriving from data sources are reliably forked and passed to both an event database for durable storage and to a real-time event query engine, often running on different machines for robustness. Event pattern queries are defined by users using a uniform \kcep query model over past, present and future events, and include using semantic predicates on domain concepts. A query manager decomposes the query into those that are required to be executed over the persistent event stream database, those that exclusively operate on current and future events, and those that require access to both. Then the real-time query engine and event database cooperatively detect patterns spanning the end-to-end event streams and return results to users. A domain knowledge-base is accessible for resolving the semantic specifications in the query, and we use a combination of semantic and CEP query engines to process these queries.

One subtlety should be recognized when performing integrated queries over real-time and archived event streams. To ensure temporal consistency, the output of any query operating on an event stream should behave as if it was fully performed on the original real-time event stream from back in time to the present, preserving the order and with no missing or duplicate events being matched. There are practical difficulties with ensuring that a fork of the incoming event stream is atomic, and happens such that every event is seen exactly once, either by the partial query operating on the real-time stream or the persistent stream.  As a result, we may have transient situations where there is a positive (missing events) or negative (duplicate events) \emph{temporal gap} at the boundary between the part of the event stream available to the archive engine and to the real-time engine. The integrated execution of queries has to be cognizant of a potential temporal gap at this boundary, and ensure that the results of the query executions are consistent. We later discuss query rewriting and execution mechanisms to guarantee this.

Our proposed \kcep model, architecture and optimizations, though motivated by Smart Grid applications, are designed as a generic system. Other potential domains include e-commerce~\cite{Asaf2006}, digital healthcare~\cite{wang:vldb:2011} and various IoT domains which feature multi-disciplinary information spaces and the need for robust analysis of streaming data.

\section{\kcep Event and Query Model}
\label{sec:eventpattern}
Allowing users to specify event query patterns with domain knowledge concepts included within them as predicates first requires that the events themselves be enhanced with semantic context about the domain~\cite{zhou2012ISWC}. We first introduce such a semantic event model followed by the \kcep query specification over them. 

\subsection{Semantic Event Model}
\begin{figure}[t]
  \centering
  \includegraphics[width=0.5\columnwidth]{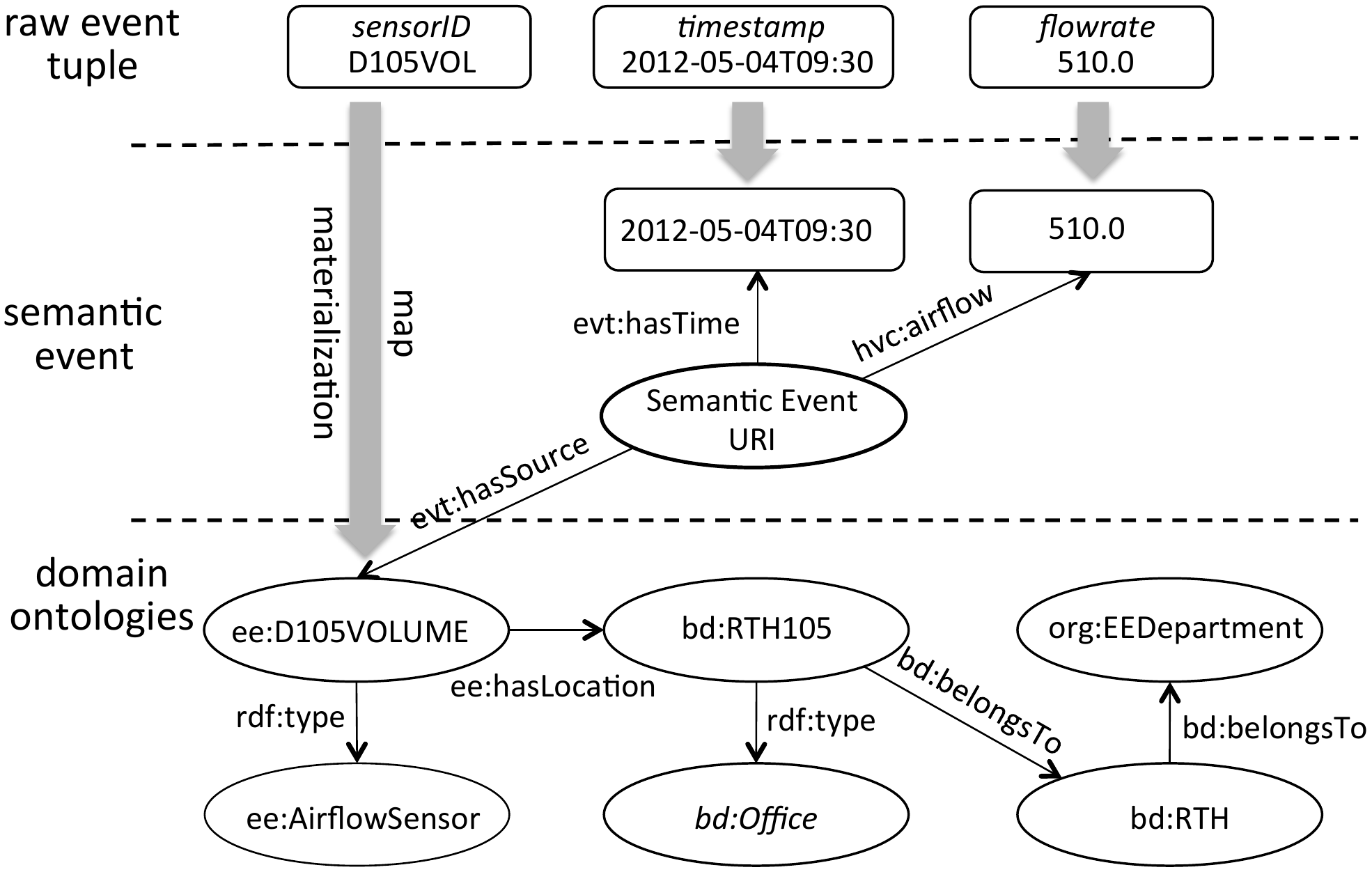}
  \caption{Semantic enrichment of \emph{AirflowReport} Event. Top row has the original raw event with attribute names and values. The bottom row shows the domain ontology that the event attribute name and value are mapped to. The middle has the semantic event that relates the \emph{flowrate} attribute name and the \emph{D105VOL} attribute value to the ontology entities \emph{hvc:airflow} and \emph{ee:D105VOLUME} respectively. Note that the semantic event also captures the relationship between the attributes in the tuple using \emph{evt:hasTime} and \emph{evt:hasSource}.}
  \label{fig:dsemantics}
\end{figure}

Traditional CEP systems treat events as syntactic data, represented in various structural formats such as relational tuples, XML, JSON and POJO~\cite{Rozsnyai2007}. We adopt a tuple-based model of a set of name-value attribute pairs for referring to each \emph{raw} syntactic event~\cite{Demers2007}, defined as: 
\begin{center} \texttt{\small Raw Event $\coloneq$ \{$\langle$name, value$\rangle\star$\}}\end{center}

To provide a higher level abstraction for user queries, we leverage ontologies specified using the Web Ontology Language (OWL)~\footnote{Web Ontology Language, \url{http://www.w3.org/TR/owl-ref}} to associate domain entities with raw events. While small-scale IoT deployments may find it adequate to normalized the raw event schema to a standard vocabulary of domain terms captured by a relational schema, using semantic ontologies helps leverage existing knowledge-bases on diverse domains developed by domain experts~\cite{zhou2012ITNG}, transparently reason over equivalent and related concepts, and include new concepts rapidly. 

The ontologies are organized in a modular fashion, allowing components from related domains to be linked together using \emph{subject--verb--object} triples that capture the relationship (\emph{verb}) between two concepts (\emph{subject, object})~\cite{zhou2012ITNG}. We enrich raw events with \emph{semantic context} by mapping attributes of their tuple to their equivalent entities in an ontology. A sample enrichment of a raw event using semantics is shown in Fig.~\ref{fig:dsemantics} where the attribute name \emph{flowrate} for the \emph{AirflowReport} event from a HVAC unit is asserted to be the same as the standard domain concept of \emph{hvc:airflow}. Rather than treat this just as a simple term-mapping exercise, we instead create a \emph{Semantic Event} that maps attribute names and values to ontology concepts, and also captures the relationship between the attributes in a tuple using verbs like \emph{evt:hasTime} and \emph{evt:hasSource}. \kcep queries are then defined over a stream of these knowledge-infused semantic events.

We distinguish semantic mappings between those associated with attribute names, which tend to be invariant or slow changing, and those mapped from attribute values, which can change rapidly from event to event. \emph{Static semantics} capture the mapping between an event schema (i.e., attribute names) and semantic concepts. For e.g., while the \emph{AirflowReport} event from one set of sensors may have an attribute name as \emph{flowrate}, other events of the same or similar type may refer to the conceptually identical \emph{flowrate} observation as \emph{airflow} or \emph{flowvolume}. A static semantic mapping relates all three of these attribute names to the standard \emph{hvc:flowrate} semantic entity defined in the HVAC ontology, having the namespace \emph{hvc}, using the \emph{owl:sameAs} relation. Hence, static semantics help address structural heterogeneity of events. 

\emph{Dynamic semantics}, on the other hand, map from the actual value of a event attribute to semantic entities, and these values can vary for each event. For e.g., in Fig.~\ref{fig:dsemantics}, the semantic event relates the value of the \emph{sensorID} attribute, \emph{D105VOL}, with a specific entity for that sensor, \emph{ee:D105VOLUME}, that is present in the domain ontology. The ontology further allows that sensor to directly be associated with the room it is present in (\emph{bd:RTH105}) and the type of variable measured by that sensor (\emph{ee:AirflowSensor}), and transitively be linked to the type of that room (\emph{bd:Office}) and the building and department it belongs to (\emph{bd:RTH} and \emph{org:EEDepartment}).

This exposes the power of the semantic event in capturing relationships between a raw event and concepts in a variety of ontologies, spanning the electrical infrastructure in the microgrid (\emph{ee:*} and \emph{hvc:*}), the buildings on campus (\emph{bd:*}), and the organizational structure (\emph{org:*}). The ontologies themselves are defined once and reused across different semantic event mappings, and existing general purpose ontologies such as on the weather~\footnote{SWEET Ontology for Earth and Environment, https://sweet.jpl.nasa.gov/} can also be linked, as we have shown before~\cite{zhou2012ITNG}.
This allows the users to define their CEP query based on well-understood domain concepts, even though they may not directly be present in the raw event but are inferred by the semantic event from the knowledge-base. So a facility manager interested in HVAC events from a certain type of physical space, say an \emph{Office} or a \emph{Meeting} room, can use the semantic event to navigate to that concept from the \emph{sensorID} attribute's value present in the \emph{AirflowReport} event. Next, we discuss the query model to specify such \kcep patterns.

\subsection{\kcep Query Model}
\label{sec:scepquery}
We propose a two-segment \kcep query model that loosely couples the query predicates over the semantic knowledge concepts and syntactic CEP predicates, and further supports querying over end-to-end archived and real-time streams. Separating out the semantic and syntactic predicates helps simplify the specification by the user, and as we see later, also allows for pipelined execution by the engine.

The general query structure, loosely based on the SPARQL semantic query language~\footnote{SPARQL Protocol and RDF Query Language (SPARQL), \url{http://www.w3.org/TR/rdf-sparql-query}}, is given by:

\noindent \\ \hspace*{4.3cm}\texttt{\small \kcep Query $\coloneq$} 
\\ \hspace*{5.2cm}\texttt{ \small PREFIX   <ontology namespaces>} .
\\ \hspace*{5.2cm}\texttt{ \small SELECT  <output event definition>} .
\\ \hspace*{5.2cm}\texttt{ \small FROM\ \ \ <input event definition>} .
\\ \hspace*{5.2cm}\texttt{ \small WITHIN <query boundary>} .
\\ \hspace*{5.2cm}\texttt{ \small WHERE\ \ [Semantic Subquery]$\star$ | [CEP Subquery]$?$
\\} 

The \texttt{PREFIX} clause defines domain ontology name\-spaces that can be referenced in both semantic and CEP subqueries. Particularly it enables qualified access to static event semantics in CEP subqueries. The \texttt{SELECT} clause \emph{projects} properties of matching events, such as attribute values and results of aggregation functions, and returns them as part of the query output event stream. The optional keyword \texttt{AS} can be used to rename event attributes in the output, and this gives syntactic control over the output event format. Output event streams resulting from one query can be further queried upon. The input event streams are identified by the keyword \texttt{FROM}. For e.g., an event placeholder \emph{e} present in the input stream \emph{airflowReport} is declared as:

\begin{center}\texttt{\small FROM (?e, airflowReport)}\end{center}

The \texttt{WITHIN} clause specifies the lower and upper temporal bounds of the events of interest, which may include events with timestamps in the past. For e.g., the default temporal boundary for traditional CEP queries that match only future events is given using the \texttt{now} keyword:

\begin{center}\texttt{\small WITHIN [now, )}\end{center}

\noindent while a time range that starts at a point in the past and spans to all current and future events may be given by:

\begin{center}\texttt{\small WITHIN [2012-05-07T09:00, )}\end{center}

\noindent The parentheses `\texttt{(}' and `\texttt{)}' indicate the start or end time is inclusive, while the square brackets `\texttt{[}' and `\texttt{]}' indicate the times are exclusive. 
This simple model extension using the \texttt{WITHIN} clause, along with its more involved runtime implementation, offers the ability to query over archived and real-time streams.

Finally, the \texttt{WHERE} clause specifies a pipeline of subqueries to filter events in the input stream. Specifically, zero or more \emph{semantic subquery} segments, combined using \texttt{AND} or \texttt{OR} keywords, specify semantic filtering constraints on events, based on both raw event tuples' attributes and values, as well as on the domain ontologies navigated to from the corresponding semantic event for the raw event. The semantic subqueries are represented using the SPARQL language, modeled as a query graph with multiple property paths~\footnote{SPARQL Property Paths, \url{http://www.w3.org/TR/sparql11-property-paths}}, that can query OWL ontologies. Each subquery operates on just one of the several event variables, if present. Support for semantic correlation across multiple event types is left as future work. 

Semantic subqueries are labeled using the keyword \texttt{PATH} as:

\begin{center}\texttt{ \small Semantic Subquery $\coloneq$ PATH <SPARQL triple patterns>}\end{center}

The \texttt{WHERE} clause can alternatively include zero or one \emph{CEP subquery}, which subsumes traditional CEP temporal and syntactical constraints over raw events. CEP query languages~\cite{Demers2007, Anicic2012, website:Siddhi} typically extend SQL relational query model with temporal operators like \emph{window} and \emph{sequence}. CEP patterns are classified as \emph{dimensional patterns} or \emph{basic patterns}, depending on whether they are related to temporal and space ordering of events, or not~\cite{Etzion2010}. 

We discretize composite CEP operators present in existing CEP query languages into unit operators, each representing a single constraint from the following categories: non-correlation, value-based correlation, and time/length-based correlation. These help capture composite CEP patterns such as \emph{basic}, \emph{threshold} and \emph{temporal} patterns, described in detail elsewhere~\cite{Etzion2010}. This also facilitates query rewriting for persisted events, discussed later. 

Specifically, we define CEP subqueries as:

\noindent \\ \hspace*{4.4cm}\texttt{ \small CEP Subquery $\coloneq$}
\\ \hspace*{5.1cm}\texttt{ \small [FILTER          <non-correlation constraint>]*}
\\ \hspace*{5.1cm}\texttt{ \small [JOIN    <value-based correlation>]*}
\\ \hspace*{5.1cm}\texttt{ \small [SEQ      <temporal order correlation>]?}
\\ \hspace*{5.1cm}\texttt{ \small [WINDOW <temporal range correlation>]?
\\} 

The non-correlation \texttt{FILTER} operator defines constraints for individual events based on attribute values. The correlation operators define constraints across multiple events based on non-temporal attributes (\texttt{JOIN}) or temporal attributes (\texttt{SEQ} and \texttt{WINDOW}).

We use the campus microgrid to illustrate common the \kcep query types of filtering, aggregation and sequence, when combined with knowledge-infused semantic predicates. These example queries are also reused in later sections. The queries use events from the \emph{airflowReport} stream that generates events on the volume of air flowing out of an HVAC unit that the given sensor ID is monitoring, and is linked to ontologies as shown in Fig.~\ref{fig:dsemantics}. For brevity in the examples below, the \texttt{FROM} clause is skipped and it implicitly refers to the \emph{airflowReport} stream. Similarly, the following ontology namespace prefixes~\cite{zhou2012ITNG} are used in the queries, and not restated below. 
\vspace{0.035in}
\\ \hspace*{4.7cm}\texttt{\small{PREFIX bd:<http://cei.usc.edu/Building.owl\#>}}
\\ \hspace*{4.7cm}\texttt{\small{PREFIX ee:<http://cei.usc.edu/Equipment.owl\#>}}
\\ \hspace*{4.7cm}\texttt{\small{PREFIX hvc:<http://cei.usc.edu/HVAC.owl\#>}}
\\ \hspace*{4.7cm}\texttt{\small{PREFIX org:<http://cei.usc.edu/Organization.owl\#>}}
\\ \hspace*{4.7cm}\texttt{\small{PREFIX evt:<http://cei.usc.edu/Event.owl\#>}}
\\ \hspace*{4.7cm}\texttt{\small{PREFIX rdf:<http://www.w3.org/1999/02/22-rdf-syntax-ns\#>}} 

\subsubsection{Example: Syntactic Filtering Query}
\label{sec:simplecep}
Our \kcep model supports plain CEP queries, a simple form of which has just filter constraints. E.g.,

\begin{query}\label{q:simplecep}
Report the sensor ID and flowrate for situations when the inbound airflow of a space exceeds $500$~cfm (cubic feet per minute).
\end{query}

\noindent \hspace*{4.5cm}\texttt{ \small SELECT ?e.sensorID, ?e.flowrate}\nopagebreak[4]
\\ \hspace*{4.5cm}\texttt{ \small FILTER (?e.flowrate > 500)\nopagebreak[4]
\\}

This syntactic CEP constraint that relies on the specific attribute name can be replaced by a more robust semantic CEP predicate that uses the static event semantics. For e.g., the \emph{flowrate} attribute is equivalent to its semantic concept, \emph{hvc:airflow}, that is captured in the HVAC ontology, but the latter would also match other airflow sensors that may use a different syntactic name:

\noindent\\ \hspace*{4.5cm}\texttt{ \small FILTER (?e.hvc:airflow > 500)
\\}

\subsubsection{Example: Syntactic Aggregation Query}
Aggregation functions such as \emph{average} and \emph{sum} can be computed over attribute values present in event streams by grouping events into (sliding or batch) moving windows, and are matched continuously over events in the stream.

\begin{query}\label{q:aggcep} 
Report the 5-minute average inbound airflow of a space when the average value is greater than $500$~cfm.
\end{query}

\noindent  \hspace*{4.5cm}\texttt{ \small SELECT AVG(?e.flowrate) > 500 AS avgrate}
\\ \hspace*{4.5cm}\texttt{ \small WINDOW (?e, sliding, 5min)
\\}

\subsubsection{Example: Syntactic Sequence Query}
CEP queries can also assert temporal ordering over events present in moving windows.

\begin{query}\label{q:seqcep}
Report the ID of the sensor when the airflow of an HVAC it is monitoring is greater than $500$~cfm, and then increases by $50$~cfm within a $5$~minute period.
\end{query}

\noindent  \hspace*{4.5cm}\texttt{ \small SELECT ?e1.sensorID}
\\ \hspace*{4.5cm}\texttt{ \small FILTER (?e1.flowrate > 500)}
\\ \hspace*{4.5cm}\texttt{ \small JOIN (?e2.sensorID = ?e1.sensorID)}
\\ \hspace*{4.5cm}\texttt{ \small JOIN (?e2.flowrate-?e1.flowrate > 50)}
\\ \hspace*{4.5cm}\texttt{ \small SEQ (?e1, ?e2)}
\\ \hspace*{4.5cm}\texttt{ \small WINDOW (?e1, ?e2, 5min)
\\} 

\subsubsection{Example: Semantic Filtering Query}
A simple semantic query includes semantic constraints over events in addition to the CEP subquery filters. The following example shows how a department's energy coordinator can extend Query~\ref{q:simplecep} with dynamic semantics. Here, the coordinator can specify the query without knowing the details of the sensor deployment within the physical spaces (i.e., \emph{Office} and \emph{EEDepartment}), and make use of the ontological inference that is enabled by our enriched semantic events and semantic subqueries.

\begin{query}\label{q:simplescep}
Report the sensor ID and flowrate when the airflow in an \emph{Office} present in the \emph{EEDepartment} exceeds $500$~cfm.
\end{query} 

\noindent  \hspace*{4.5cm}\texttt{ \small SELECT ?e.sensorID}
\\ \hspace*{4.5cm}\texttt{ \small FILTER (?e.hvc:airflow > 500) }
\\  \hspace*{4.5cm}\texttt{ \small PATH \{?e evt:hasSource ?src .}
\\ \hspace*{5.4cm}\texttt{ \small ?src rdf:type ee:AirflowSensor .}
\\ \hspace*{5.4cm}\texttt{ \small ?src bd:hasLocation ?loc .}
\\ \hspace*{5.4cm}\texttt{ \small ?loc rdf:type bd:Office .}
\\ \hspace*{5.4cm}\texttt{ \small ?loc bd:belongsTo org:EEDepartment\}
\\} 

To illustrate the intuitiveness of \kcep further, we can leverage a domain concept, \emph{GreenOfficeAirflow}, defined in the HVAC ontology with a value that meets the airflow upper bound for a sustainable office space, instead of using a static $500$~cfm value in the query:
\begin{query}\label{q:simplescep2}
Match a sensor ID when the airflow in an \emph{Office} room present in the \emph{EEDepartment} exceeds the \emph{GreenOfficeAirflow} value.
\end{query} 

\noindent \\ \emph{ \hspace*{4.5cm}\texttt{ \small SELECT ?e.sensorID}
\\ \hspace*{4.5cm}\texttt{ \small PATH \{?e evt:hasSource ?src .}
\\ \hspace*{5.4cm}\texttt{ \small ?src rdf:type ee:AirflowSensor .}
\\ \hspace*{5.4cm}\texttt{ \small ?src bd:hasLocation ?loc .}
\\ \hspace*{5.4cm}\texttt{ \small ?loc rdf:type bd:Office .}
\\ \hspace*{5.4cm}\texttt{ \small ?loc bd:belongsTo org:EEDepartment .}}
\\ \hspace*{5.4cm}\texttt{ \small ?e hvc:airflow ?rate .}
\\ \hspace*{5.4cm}\texttt{ \small hvc:GreenOfficeAirflow hvc:hasValue ?gaf .}
\\ \hspace*{5.4cm}\texttt{ \small FILTER (?rate > ?gaf)\}
\\}

In the above query, we do not require the CEP \texttt{FILTER} subquery from Query~\ref{q:simplescep} while we append additional SPARQL predicates to the original (italized) semantic subquery.

\subsubsection{Example: Semantic Aggregation Query}
This query incorporates an aggregation function into the \texttt{SELECT} clause, and the function is applied over events that match both the semantic and the CEP constraints. It reflects a combination of the CEP subquery from Query~\ref{q:aggcep} that has the \texttt{AVG} function with the semantic subquery from Query~\ref{q:simplescep}.

\begin{query}\label{q:aggscep}
Report the $5$-minute average inbound airflow of an \emph{Office} that is present in the \emph{EEDepartment}, when the average value is greater than $500$~cfm.
\end{query} 

\noindent  \hspace*{4.5cm}\texttt{ \small SELECT AVG(?e.flowrate) > 500 AS avgrate}
\\ \hspace*{4.5cm}\texttt{ \small WINDOW (?e, sliding, 5min) }
\\  \hspace*{4.5cm}\texttt{ \small PATH \{?e evt:hasSource ?src .}
\\ \hspace*{5.4cm}\texttt{ \small ?src rdf:type ee:AirflowSensor .}
\\ \hspace*{5.4cm}\texttt{ \small ?src bd:hasLocation ?loc .}
\\ \hspace*{5.4cm}\texttt{ \small ?loc rdf:type bd:Office .}
\\ \hspace*{5.4cm}\texttt{ \small ?loc bd:belongsTo org:EEDepartment\} }

\subsubsection{Example: Semantic Sequence Query}
A semantic sequence query is one that specifies semantic constraints over events while also applying sequence ordering as CEP constraints. As shown in the example below, the CEP subquery (italized) is identical to Query~\ref{q:seqcep} which uses \texttt{SEQ} to order events in \emph{$5$~min} sliding windows. In addition, two semantic subqueries, one for each component event \emph{e1} and \emph{e2} in the sequence, are specified to constrain them to meeting rooms. 

\begin{query}\label{q:seqscep} Report the ID of the sensor when the airflow of a \emph{MeetingRoom} it is monitoring is initially greater than $500$~cfm, and then increases by $50$~cfm within a $5$~minute period.
\end{query}

\noindent \\ \emph{ \hspace*{3.5cm}\texttt{ \small SELECT ?e1.sensorID \ \  FILTER (?e1.flowrate > 500)}
\\ \hspace*{3.5cm}\texttt{ \small JOIN (?e2.sensorID = ?e1.sensorID) \ \  JOIN (?e2.flowrate-?e1.flowrate > 50)}
\\ \hspace*{3.5cm}\texttt{ \small SEQ (?e1, ?e2) \ \  WINDOW (?e1, ?e2, 5min) } }
\\ \hspace*{3.5cm}\texttt{ \small PATH \{?e1 evt:hasSource ?src1 .} \nopagebreak[4]
\\ \hspace*{4.4cm}\texttt{ \small ?src1 rdf:type ee:AirflowSensor .} \nopagebreak[4]
\\ \hspace*{4.4cm}\texttt{ \small ?src1 bd:hasLocation ?loc1 .} \nopagebreak[4]
\\ \hspace*{4.4cm}\texttt{ \small ?loc1 rdf:type bd:MeetingRoom .} 
\\ \hspace*{4.4cm}\texttt{ \small ?e2 evt:hasSource ?src2 .} \nopagebreak[4]
\\ \hspace*{4.4cm}\texttt{ \small ?src2 rdf:type ee:AirflowSensor .} \nopagebreak[4]
\\ \hspace*{4.4cm}\texttt{ \small ?src2 bd:hasLocation ?loc2 .} \nopagebreak[4]
\\ \hspace*{4.4cm}\texttt{ \small ?loc2 rdf:type bd:MeetingRoom \} } \\

The above scenarios illustrate the intuitive nature of \kcep queries specified over a single sensor in a campus microgrid IoT environment. The knowledge infusion enables a higher level of abstraction when defining queries, which makes use the semantic ontologies and precludes the need for the end user to know fine-grained details of the event format or the deployment model, both of which could change often.

\section{Processing \kcep Queries over  Real-time Streams}
\label{sec:streamscep} 
In this section, we describe approaches to detect \kcep event patterns, that include both syntactic and semantic CEP subqueries, exclusively from \emph{real-time} event streams. In later sections, we extend this to \kcep queries over archived streams, and over end-to-end streams that span archive and real-time.

General purpose semantic SPARQL query and inferencing can be complex, and are time consuming even for in-memory processing~\cite{David2012}. Since our \kcep query model is restricted to static and dynamic semantics that have enriched the raw events, and we support syntactic CEP queries (that can be processed rapidly) in addition to semantic ones, this offers opportunities to make the runtime execution of \kcep queries by our query processing engine more efficient. Specifically, we explore optimizations at \emph{query compile time}, when the query is submitted by the user, and at \emph{query runtime}, when the events arrive for processing. 

\subsection{Compile-time Query Optimization}
\label{sec:compile}
\kcep query predicates that are defined only on static event semantics, i.e. knowledge concepts associated with event schemas or static semantic knowledge-base, can be evaluated offline since their result will not change based on event values at runtime. Query \ref{q:simplescep2} from the earlier example falls in this category. We propose three query optimizations that can be performed at compile-time to leverage this.

\textbf{Semantic pruning and migration.} This reduces the complexity of semantic subqueries. In the pruning step, SPARQL property paths within semantic subqueries which originate from \emph{ontology constants}, such as classes and instances, are executed in advance and their results replaced within the query clauses. For e.g., in Query~\ref{q:simplescep2} the property path from \texttt{hvc:GreenOfficeAirflow} concept can be replaced by its literal value \texttt{500} in the subquery. Here, the expectation is that such constants in the ontology will rarely change, and if they do, the queries affected by this optimization will be refreshed to use the new values.

\textbf{Transformation optimization.} This leverages the fact that pure CEP can typically be executed much more efficiently than semantic query predicates. Here, we rewrite one-hop semantic property paths originating from an event variable into more efficient CEP subquery clauses. For e.g., the last three clauses of the semantic subquery of Query~\ref{q:simplescep2} can be completely transformed to a CEP \texttt{FILTER} subquery, similar to Query~\ref{q:simplecep}, once the semantic pruning has replaced \texttt{hvc:GreenOfficeAirflow} with the value $500$. 

\textbf{Semantic normalization.} This eliminates static semantics from CEP subqueries. Raw event attributes are standardized in a specific domain. Equivalent event attributes in the raw event streams are mapped to conform to the standard. Event attributes present in CEP subqueries are also normalized to standard terms at compile time so that semantic inference for CEP subqueries are not repeated for each arriving event. For e.g., the same concept, \emph{airflow}, has alternate attribute names like \emph{flowrate} and \emph{airvolume} in the \emph{AirflowReport} streams arriving from different sources, any of which names may be used in a CEP subquery. A static semantic inferencing using the domain ontologies can normalize them offline to just one of these equivalent terms. 

These compile-time optimizations reduce the number and complexity of semantic predicates in \kcep queries, thus avoiding repetitive and costly semantic reasoning and evaluations at runtime.

\subsection{Runtime Query Optimization}
\label{sec:runtime}
We adopt an asynchronous pipelined architecture to process \kcep query segments at runtime. As shown earlier in Fig.~\ref{fig:dsemantics}, raw event tuples that arrive on streams (top row) are annotated and linked with static domain ontologies (bottom row) to form semantically-enriched events (middle row).  The semantic events are passed to a \emph{semantic filter module} which evaluates semantic subqueries. Events that satisfy these semantic constraints are further allowed to pass to a CEP engine that evaluates the CEP subqueries. Our proposed optimizations focus on runtime semantic subquery processing; traditional CEP engines are as such efficient for CEP subqueries. 

A na\"{i}ve approach is to evaluate the semantic subqueries for every new event that arrives, using the domain ontologies. However, evaluating a SPARQL query requires costly inferencing and self-join operations over the knowledge base -- typically, a SPARQL query with a single property path of length $n$ requires $(n-1)$ self-joins over the ontology~\cite{Zhang05ontologyquery}. Even using an in-memory semantic query engine, we observe evaluation throughputs flatten at $\sim$80 events per second in our experiments (\S~\ref{sec:experiments}), while CEP engines can process syntactic queries at $1000$'s of events per second. We propose two runtime semantic subquery optimizations to mitigate this.

\subsubsection{Event Buffering}
\label{sec:buffer}
Buffering data streams for lazy query processing has been proposed before, such as for XML stream querying~\cite{Todd2004} and for CEP processing in T-Rex~\cite{Cugola2012}. Those approaches were studied for syntactic data and queries, and T-Rex's lazy processing using buffers is actually less effective than an automata-based eager evaluation for CEP sequence patterns. However, for a semantic subquery in \kcep, the time to evaluate it for one event is comparable to that for evaluating it over a small set of events, due to the high static overhead of semantic query processing. Hence, we posit that event buffering will be effective.

Rather than evaluate semantic subqueries for each event upon arrival, we instead buffer events that arrive within a (configurable) time interval and perform the query collectively on this batch of events. The obvious side-effect of this is the introduction of a delay in pattern detection that, in the worst case, equals the \emph{buffer interval duration}. So the interval length should be small enough that the user application can tolerate the delay. 

It is intuitive that as long as the query processing time for a batch of events is less than the buffer interval duration, the query throughput can keep up with the input event rate. However, as shown in our experiments (\S~\ref{sec:experiments}), this is subject to the input event rate being adequately large. When the input rate is small, the batch may have just one event -- degenerating to the baseline of processing one event at a time, and now with a additional delay. On the other hand, if the input rate is very high, the time to perform the semantic query for the batch of events can exceed the buffer window, causing the query throughput to fall below the input rate. Thus, the choice of buffer interval is a careful trade-off between query latency and maximum input throughput. 

\subsubsection{Semantic Caching} 
\begin{figure}[t]
  \centering 
  \includegraphics[width=0.55\columnwidth]{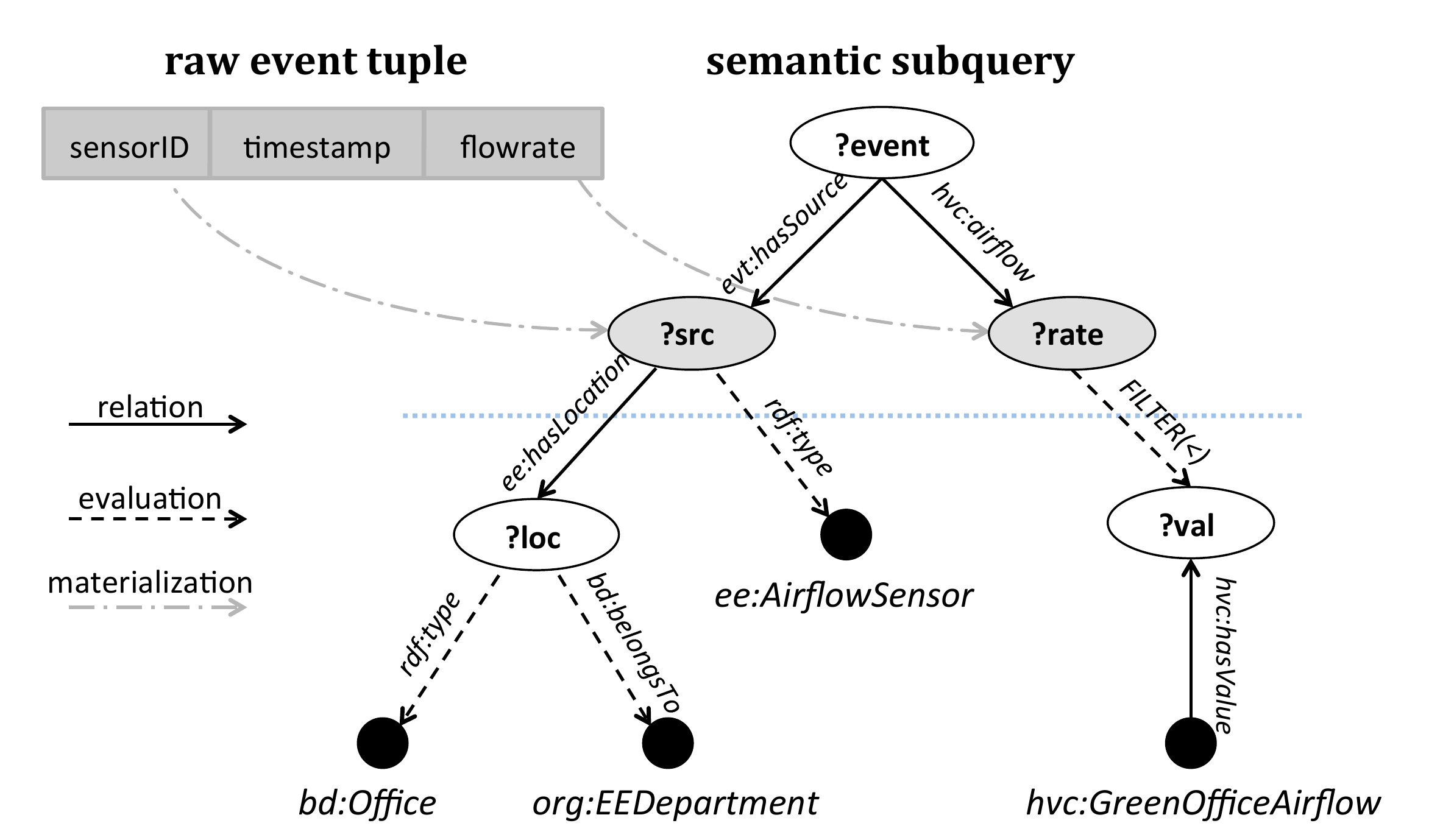}
  \caption{Query tree graph for the property paths of the semantic subquery in Query \ref{q:simplescep2}}
  \label{fig:qgraph}
\vspace{-0.2in} 
\end{figure}

Yet another optimization we propose is caching semantic subquery results, similar to the caching mechanisms employed to speed up memory, database and web data accesses. The key intuition for semantic subquery caching is that multiple event tuples may share the same value for some attribute, and the evaluation results for queries specified on those events with same attribute values can be reused. 

Consider the semantic event that is enriched and materialized from a raw event tuple as a \emph{event tree graph} that is rooted at the event URI node with edges linking to property nodes, as seen in Fig.~\ref{fig:dsemantics}.

\textbf{Definition 1.} \emph{The \textbf{event root properties} of a semantic event are the ontology properties that are directly materialized from the raw tuple's attribute values.}

For example, in Fig.~\ref{fig:dsemantics}, the event root properties are \texttt{`ee:D105VOLUME'}, \texttt{`2012-05-04T09:30'} and \texttt{`510.0'}. 

On the other hand, a SPARQL semantic subquery which consists of multiple property paths can also be modeled as a \emph{query tree graph} whose root nodes are event variables present in the query (e.g., \emph{?e} in Query~\ref{q:simplescep2}), the inner nodes are property variables (e.g., \emph{?src} and \emph{?loc}) and leaf nodes are constants including literals, ontology classes and instances (e.g., \emph{bd:Office} and \emph{hvc:GreenOfficeAirflow})~\cite{zhou2012ISWC}. This is shown in Fig.~\ref{fig:qgraph}. Evaluating a semantic subquery over an event essentially consists of checking if the event tree leads to the same set of leaf nodes as the query tree. Specifically we have:

\textbf{Definition 2.} \emph{The \textbf{query root properties} of a semantic event for a given semantic subquery are its \emph{event root properties} which are evaluated in the query.}

Obviously events that have the same \textit{query root properties} share the same subquery
evaluation result. For e.g., in Query~\ref{q:simplescep2}, different airflow measurement events with the same
\emph{sensorID} will have the same originating location type, and thus return the same
boolean result for the semantic query: \emph{Is the location of type \emph{Office} and does it belong to the \emph{EEDepartment}?} 

Based on this reasoning, we develop a caching technique where a cache is maintained for each semantic subquery that is submitted by the user. The key to a cache entry is formed using a canonical combination of the event root properties of an incoming event, and the value present in this entry is the boolean result of the evaluation of the semantic subquery on the event. The cache is implemented as a linked hashmap with a fixed number of entries, and we use a Least Recently Used (LRU) algorithm for cache eviction. 

Both event buffering and semantic caching are applied together during real-time query processing, and their performance results reported in \S~\ref{sec:experiments}.

\section{Processing \kcep Queries over Persistent Streams}
\label{sec:dbprocessing}
Processing \kcep queries over end-to-end event streams requires querying over persistent event streams to supplement the prior discussion on evaluating the query model for real-time event streams. Here, events arriving on a stream are also forked and persisted to a database that can be queried. We propose several incrementally more sophisticated solutions to achieve this design. 

\subsection{Event Replay}
In the most simple and direct approach, all events which occurred in the past and fall within the time range of the \texttt{WITHIN} clause of the \kcep query, are extracted from the event database and then replayed to the real-time query engine, preceding the events present in the real-time stream. This requires minimal query processing capability from the event archive database -- a relational database or even a key-value store indexed by the event timestamp will suffice. The real-time engine is fully responsible for performing the \kcep query, using the techniques discussed above, on these replayed and real-time events appearing on a unified event stream. This approach is used by existing CEP systems like Oracle Complex Event Processing as it needs limited additional tooling to implement.

However, the performance of a na\"{i}ve event replay falls short when the queries operate over a long history of high-rate events as that would force more events to be materialized from the event database. Then the performance, as
measured by the latency and throughput of matched patterns, depends on the ability of the real-time engine to manage a large burst of archived events. While syntactic CEP queries can typically be processed with high throughput, real-time semantic query evaluation is much more expensive. Hence, this approach may prove infeasible when the historical time range of the \kcep queries is large.

\subsection{Plain Query Rewriting}
\label{sec:plainrewrting}
\begin{figure}[t]
  \centering
  \setlength{\abovecaptionskip}{11pt plus 0pt minus 0pt}
  \includegraphics[width=0.5\columnwidth]{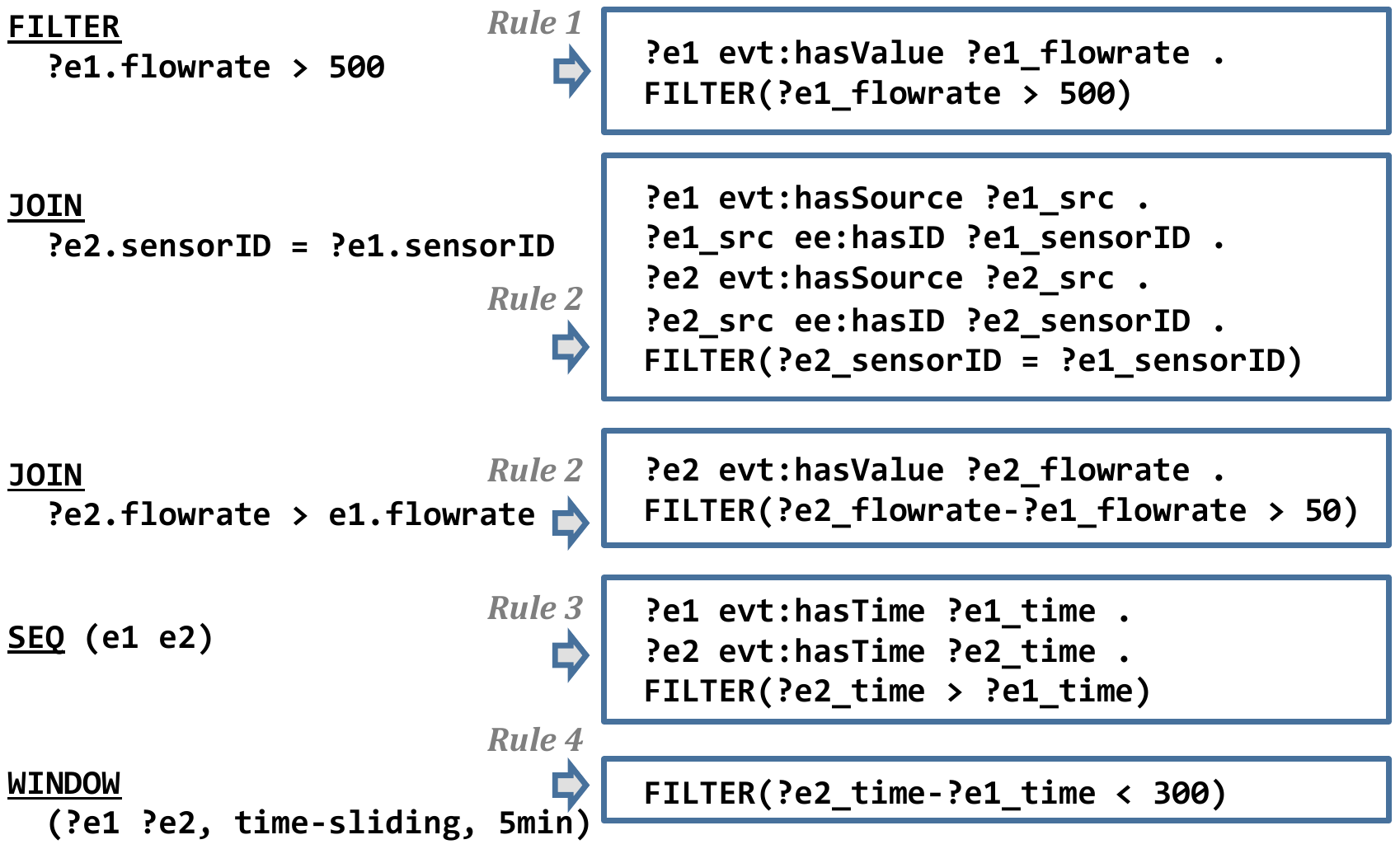}
  \caption{Rule-based Query Rewriting from \kcep to SPARQL for Query~\ref{q:seqcep}.}
  \label{fig:rewriting}
\end{figure}

An alternative approach to process the persistent event streams is to push the \kcep queries to the event archive database. The use of Semantic Web ontologies as knowledge-base and for knowledge-infusion means that the event database needs to be an RDF Triple store and support SPARQL. Since the semantic subqueries already conform to SPARQL, only the syntactic CEP subqueries in the \kcep model need to be transformed to native SPARQL. This can be achieved using rule-based mappings from CEP clauses to SPARQL property paths that can be evaluated on the event database. We next describe rewriting rules for the basic CEP operators of the \kcep model that can then be generalized to combinations of these operators, with illustrative examples provided in Fig.~\ref{fig:rewriting}.

A CEP subquery may consist of multiple non-correlation constraints defined using the \texttt{FILTER} operator, each of which can be rewritten in SPARQL using: 

\begin{rerule}[FILTER]\label{r:filter}A CEP \texttt{FILTER} constraint on an event attribute is translated into a property path in the target SPARQL query which specifies a SPARQL \texttt{FILTER} to evaluate the same value constraint over the mapped ontology property.\end{rerule}

Similarly, it is straight-forward to rewrite CEP \texttt{JOIN} clauses that specify value-based correlation constraints into SPARQL predicates using: 
\begin{rerule}[JOIN]\label{r:rel} A CEP \texttt{JOIN} constraint over a set of raw event attributes is translated into property paths that specify a SPARQL \texttt{FILTER} to evaluate the same value constraints between the mapped ontology properties.\end{rerule}

Time-based correlation operators use rules to explicitly represent the abstract constraints using semantic time properties of events stored in the archive. For e.g., the following rule transforms \texttt{SEQ} expressions in the CEP subquery to SPARQL triple patterns: 

\begin{rerule}[SEQ]\label{r:seq} A CEP \texttt{SEQ} clause is translated into a set of property paths which use SPARQL \texttt{FILTER}s to compare the timestamps of the mapped semantic events for sequence ordering.\end{rerule} 
  
\texttt{WINDOW} constraints as shown in the CEP subquery for Query~\ref{q:seqscep} can also be interpreted as timestamp comparisons between component events. Let $?e_{first}$ be the first event and $?e_{last}$ be the last event in the window, and $w$ be the window width. The moving time window is enforced using the following rule:

\begin{rerule}[WINDOW]\label{r:windowseq}A multi-variant time \texttt{WINDOW} of window width `$w$' is translated into property paths that use a SPARQL \texttt{FILTER} to evaluate the condition that $\big( timestamp(?e_{last})-timestamp(?e_{first}) \big) \le w$.\end{rerule} 

Lastly, the replayed event stream has to be returned in the same temporal order as the original CEP event stream to assure a consistent result stream. Hence this ordering guarantee must be maintained on the results from the SPARQL query:

\begin{rerule}[Order]\label{r:orderseq}Add an \texttt{ORDER BY} clause to the target SPARQL query to sort the result by the logical timestamp of the resulting events.\end{rerule}

Fig.~\ref{fig:rewriting} shows the CEP subquery of Query \ref{q:seqcep} being rewriting into SPARQL using these rules. The query rewriting approach heavily depends on the database used to persist the archived events. For e.g., since the SPARQL query language does not explicitly support a sliding window operator, it is non-trivial to write a single-variant \texttt{WINDOW} expression such as the aggregation window in Query~\ref{q:aggcep} into a RDF database query. An optional approach is to perform the query over every possible window in the archive. Also, additional rules will need to be developed for other new CEP subquery operators based on their specific definitions.

\subsection{Hybrid of Replay and Rewriting}
Event replay and direct query rewriting process different \kcep query clauses with varying efficiencies. In particular, query rewriting may benefit from batch evaluation of semantic clauses in the database since the target SPARQL query is executed just once on the past events, rather than once per event or batch (with the optimization). However, certain CEP subquery clauses can introduce severe overheads when executed on the entire archive dataset. Some of these clauses can actually be executed more efficiently by the real-time query engine through replay. 

Using this intuition, we examine the query operators described in \S~\ref{sec:eventpattern} for the relative benefits of these two contrasting approaches. We would like non-correlation constraints that evaluate events uniformly, such as semantic subqueries and CEP \texttt{FILTER} constraints, to be processed in the database. Specifically, semantic subquery evaluation has a high static overhead introduced by inferencing over the knowledge-base. This can be mitigated when the subquery is evaluated just once for all archived events at one go. On the other hand, value and time based correlation operators should be executed in the real-time query engine. For e.g., \texttt{SEQ} constraints rewritten into SPARQL can be very expensive when evaluated on the archived events due to excessive and unnecessary joins over the time property that considers all events stored in the database. 

Given the variable performance benefits of the different query engines for evaluating \kcep queries, our prototype implementation uses a hybrid approach that leverages this arbitrage, and partitions the clauses based on the strategy that can be more efficiently evaluated by the real-time engine or the RDF database. We perform partial query rewrites, enforcing only some rules, and use the resulting events as a partial result stream on which the \kcep queries are further applied by the real-time engine. Specifically, we rewrite all query clauses, except for those that perform correlation, into SPARQL. This partial SPARQL query is executed efficiently by the RDF database to materialize a pre-filtered archived event stream that is replayed and evaluated efficiently by a CEP engine (without requiring semantic support) to complete the \kcep pattern match.  

This hybrid approach executes costly event correlations within moving windows in the real-time query engine rather than over all events in the database, as done by the plain rewriting. This also has benefits over the replay. One, the SPARQL pre-filtering produces fewer events for replay, and two, more importantly, the expensive semantic subqueries are processed in batch in the database rather than evaluated per event in the replay. 

\section{Managing Temporal Gaps in End-to-end Query Processing}
\label{sec:integrated}

\begin{figure*}[t]
\centering
  \subfloat[Zero Gap Stream]{\label{fig:zeroGap}\includegraphics[width=0.34\textwidth]{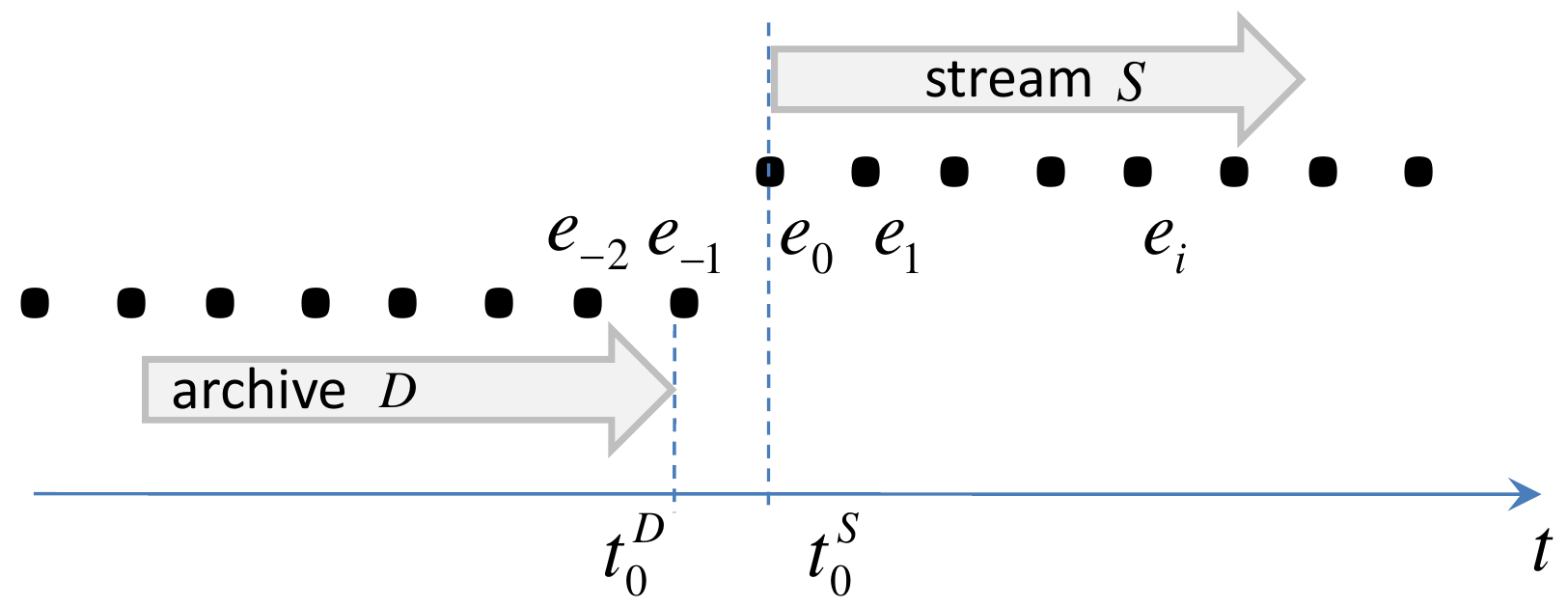}}
  \subfloat[Negative Gap Stream]{\label{fig:negGap}\includegraphics[width=0.32\textwidth]{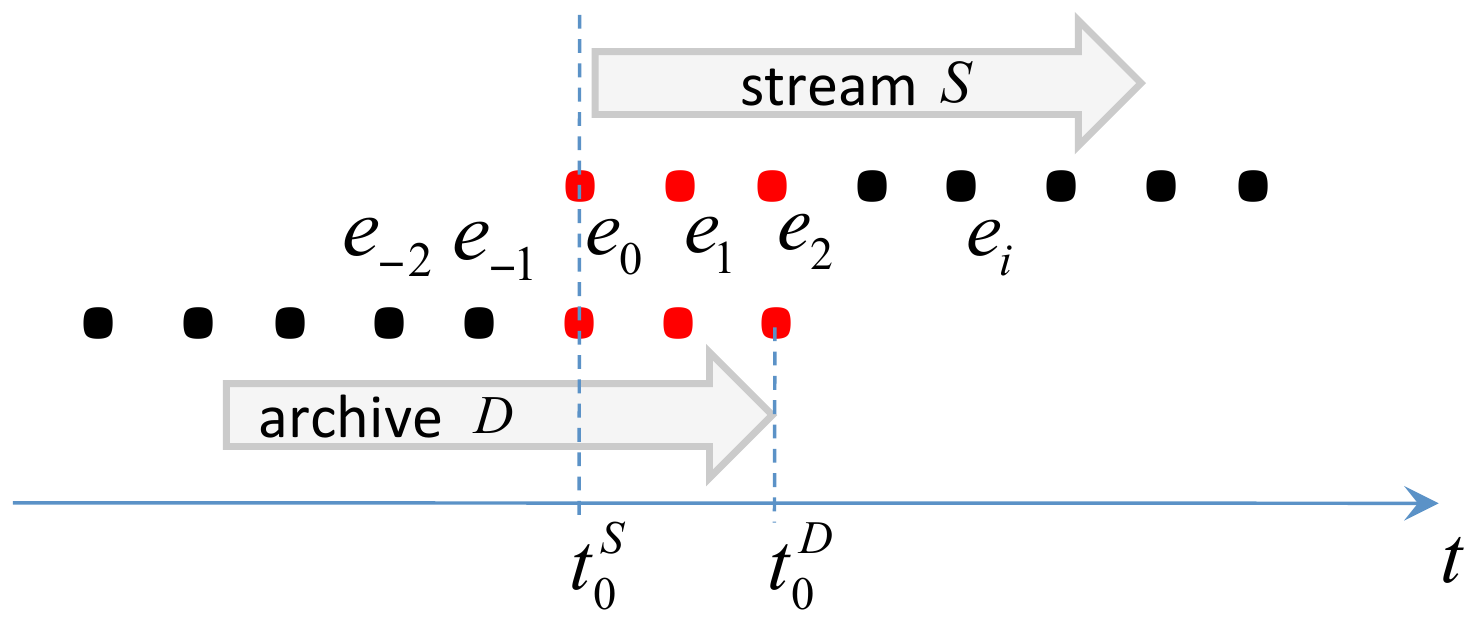}}
  \subfloat[Positive Gap Stream]{\label{fig:posGap}\includegraphics[width=0.31\textwidth]{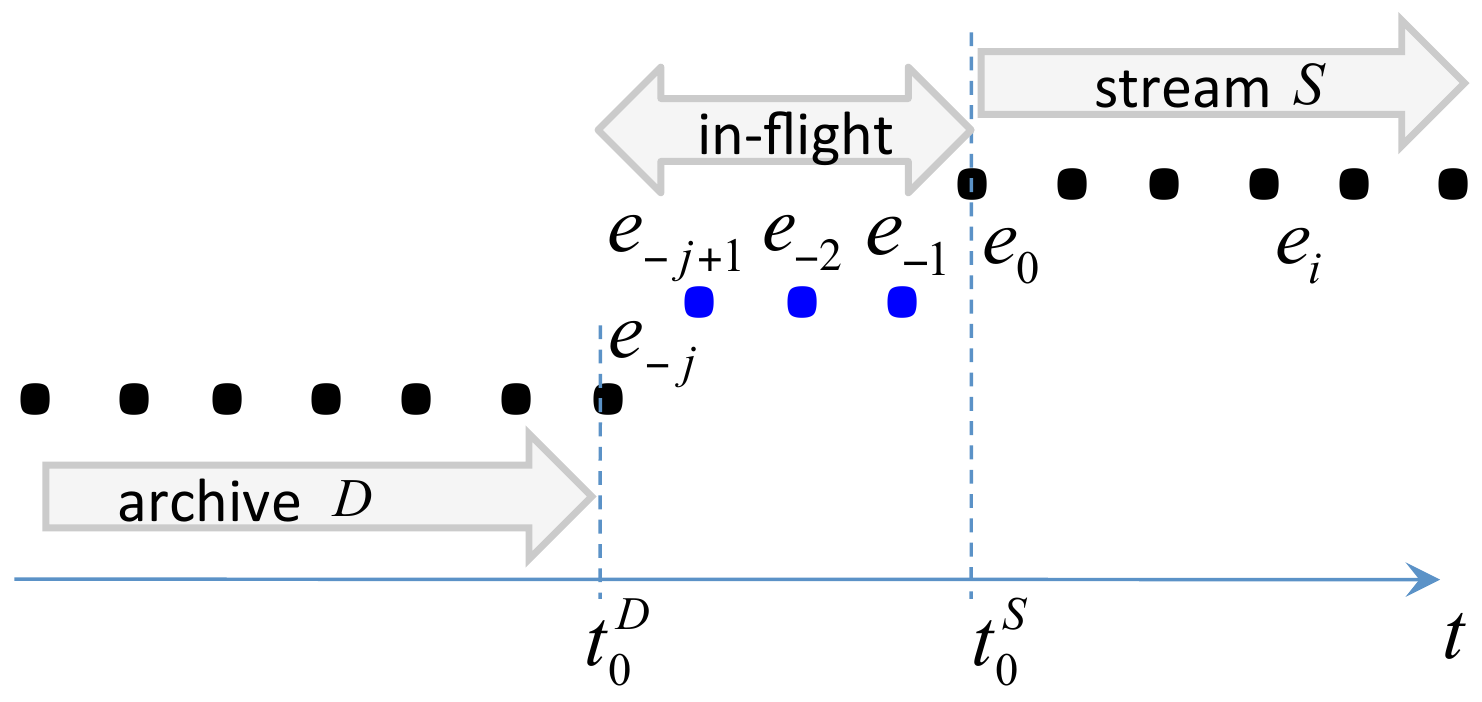}}
\caption{Configurations of a logic event stream F where events may be available only to either of the engines (a), both of them (b) or neither (c). X axis shows time. Top and bottom dots are events available to real-time and archive query engines, respectively. Red dots are duplicate events present in both engines while blue dots are in-flight events present in neither of the two engines. Vertical dotted lines are real-time and archive event stream boundaries. }
\label{fig:flow}
\end{figure*} 

The proposed \kcep query processing approaches operate over real-time and archived event streams independently, though using the same query model. In addition, we need to plan the execution of these queries performed across historic and future events and ensure the results returned to the user are in-order, \emph{and without duplicate or missing matches}. This consistency is required even when it cannot be guaranteed that \emph{an event is atomically present in either the real-time or the archived stream} -- possible hardware glitches and software delays means that an event in the input stream cannot be forked such that events seen by the real-time engine are atomically persisted into the archive database. As a result, a single logical \kcep that is being executed in a hybrid model may not see a contiguous and unique event stream that seamlessly spans the semantic database and the CEP engine when executing sequentially across these two engines. The lack of native transactional execution capability across two query environments poses unique consistency challenges when we consider queries whose temporal clause's range spans the boundary between archived and real-time events.

We can model this lack of atomicity and contiguity of the stream spanning the two engines as either a negative temporal gap (overlap) or a positive temporal gap (discontinuity) in the event stream. Fig.~\ref{fig:flow} shows possible physical presence of an end-to-end logic event stream $F$ based on the temporal gap between events available to the real-time and the archive query engines. In an ideal scenario (Fig.~\ref{fig:zeroGap}), events are atomically present in either the database stream $D$ or the real-time stream $S$, and indicates a \emph{zero} gap. There may be situations where events just stored in the archive are simultaneously available to the real-time stream too (e.g., events $e_0$, $e_1$ and $e_2$ in Fig.~\ref{fig:negGap}), and such a duplicate situation indicates a \emph{negative gap}, i.e.,  an overlap where events are potentially processed twice, once each by the real-time and the archive engines. Alternatively, in-flight events that were just seen by the real-time stream but have not yet been stored in the persistent store (e.g., events $e_{-1}$, $e_{-2}$ and $e_{-j+1}$ in Fig.~\ref{fig:posGap}) would indicate a \emph{positive gap}, and a query submitted at that instant would not see these events.

Note that we only consider a contiguous temporal gap at a single boundary location, typically present at the point where the event is forked between the two engines, and not multiple gaps at arbitrary locations in the stream. Also, such a temporal gap indicates a transient situation for individual events at the boundary that has to be managed, and an event is neither permanently lost not does it have a duplicate within the same (archive or real-time) engine.

Without loss of generality, consider a \kcep query $Q$ which operates on the logical event stream $F$. $S \subseteq F$ is the real-time stream at current time $t_0^{S}$ while $D \subseteq F$ represents the archived stream at the same time instant. Let the timestamp of the latest event available in $D$ at time $t_0^{S}$ be $t_0^{D}$. Let the first event observed by the real-time engine after $t_0^{S}$ be $e_0$, the second event be $e_1$, and so on, and the events before $e_0$ be $e_{-1}, e_{-2}, \ldots$. Due to the way stream $F$ is forked to $D$ and $S$, it is possible that (i) $S \oplus D = F$, (ii) $S  \cap D \ne \varnothing$, or (iii) $S \cup D \subsetneqq F$, each of these being mutually exclusive. These refer to zero, negative and positive gaps in the stream, and the objective is to compute a consistent query result $R$ under these conditions. 

\subsection{Query Plan for Zero Gap Streams}

In an ideal zero time gap situation, events after being visible to the real-time query engine instantaneously reach the persistent database. Integrated query plans for different \kcep queries on such streams are discussed first.

Consider a query $Q$ \emph{without} a \texttt{WINDOW} clause (e.g., Query~\ref{q:simplescep}). At time $t_0^{S}$ we apply $Q$ simultaneously to the real-time stream $S$ and the persistent stream $D$, respectively. Let the subset of patterns detected on $S$ be $R_{S}$ and patterns detected on $D$ be $R_{D}$. The integrated query result is simply $R=R_{S} \cup R_{D}$, and it guarantees no duplicate or missing patterns.

For a \kcep query $Q$ \emph{with} a \texttt{WINDOW} clause $W_{Q}$ of length $W_{Q}^{length}$, applying $Q$ on $S$ and $D$ retrieves patterns $R_{S}$ and $R_{D}$ which only contain events observed on $S$ or $D$, respectively. However, valid patterns that require component events from both real-time and archived streams are missed in $R_{S}\cup R_{D}$ since the time window can span across them. Let the missing pattern set be $R_{boundary}$. We define the starting and ending times for boundary windows on the $D$ and the $S$ streams respectively as follows. This is also visually shown in Fig.~\ref{fig:zeroGapWindow}, where the top row shows the events visible to the real-time engine and the bottom row has events present in the archive, with X axis being wall-clock time and the boundaries windows labeled at the top.

\begin{center}$W_{boundary}^{D} = (t_0^{S}-W_{Q}^{length},\  t_0^{S})$ \\
$W_{boundary}^{S} = [t_0^{S},\  t_0^{S}+W_{Q}^{length})$
\end{center}

Let $r_{boundary}\in R_{boundary}$ be the result events that are missing from $R_{S}\cup R_{D}$, and $C = \left\{c_{i}\ |\ i=0, ... n, n>0\right\}$ be the input events that contributed to $r_{boundary}$, in time order. Let $c_{i}^{timestamp}$ be the timestamp of event $c_i$. A result event $r_{boundary}$ must satisfy these necessary and sufficient conditions:

\begin{center}$c_0^{timestamp}\in W_{boundary}^{D}$\\
$c_n^{timestamp}\in W_{boundary}^{S}$\end{center}

Given this, we modify the query plan as follows to get a consistent result. Firstly query $Q$ is extended to $Q'$ by adding query clauses that represent the above temporal constraints over events $c_0$ and $c_n$. From the time $t_0^{S}$ when the query is applied, the system waits for a time period of $W_{boundary}^{S}$ to ensure all events in the boundary window $W_{boundary}^{S}$ have reached the archive. It then lazily executes $Q'$ over $D$ to retrieve $R_{boundary}$. The integrated query results are then given by $R = R_{S} \cup R_{D} \cup R_{boundary}$, ordered by the timestamp of the last component event in the query pattern. 

\subsection{Query Plan for Non-Zero Gap Streams}
\label{sec:nonzerogap}
\label{sec:zerogap}
\begin{figure*}[t]
\centering
  \subfloat[Zero Gap Scenario]{\label{fig:zeroGapWindow}\includegraphics[width=0.325\textwidth]{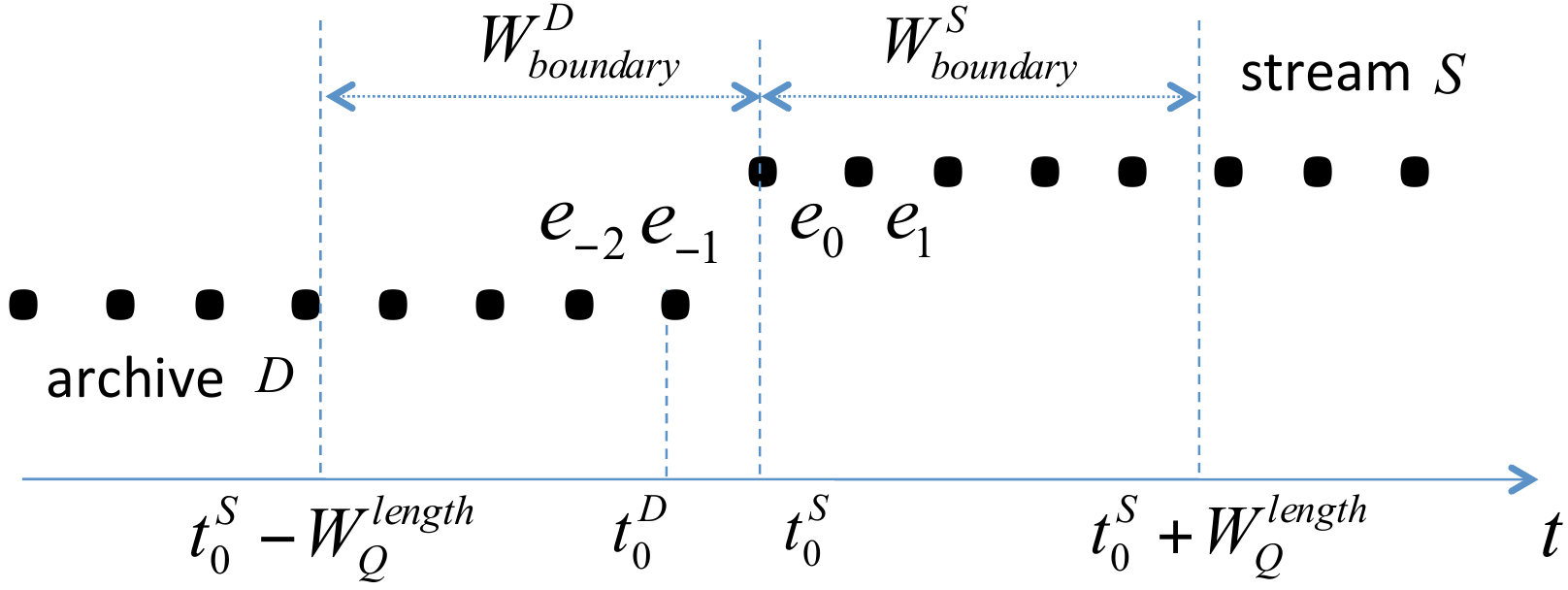}}
  \subfloat[Negative Gap Scenario]{\label{fig:negGapquery}\includegraphics[width=0.31\textwidth]{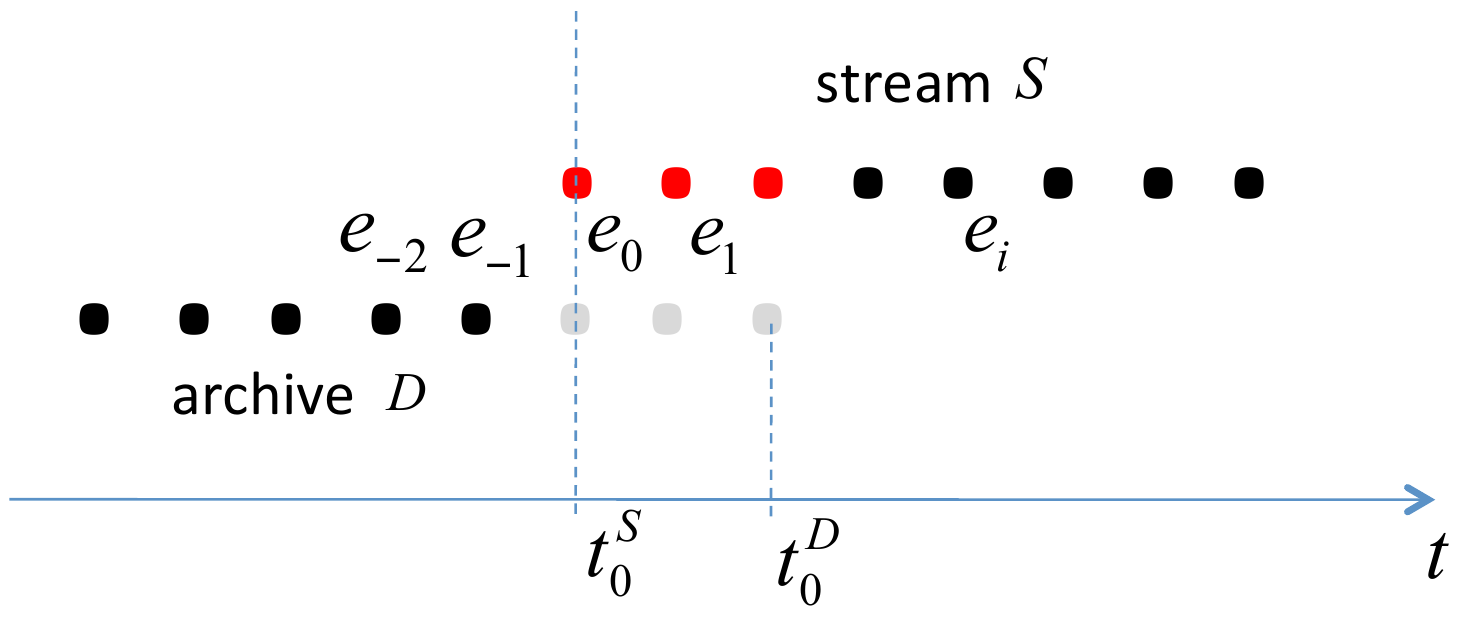}}
  \subfloat[Failure recovery with a positive gap.]{\label{fig:ha}\includegraphics[width=0.31\textwidth]{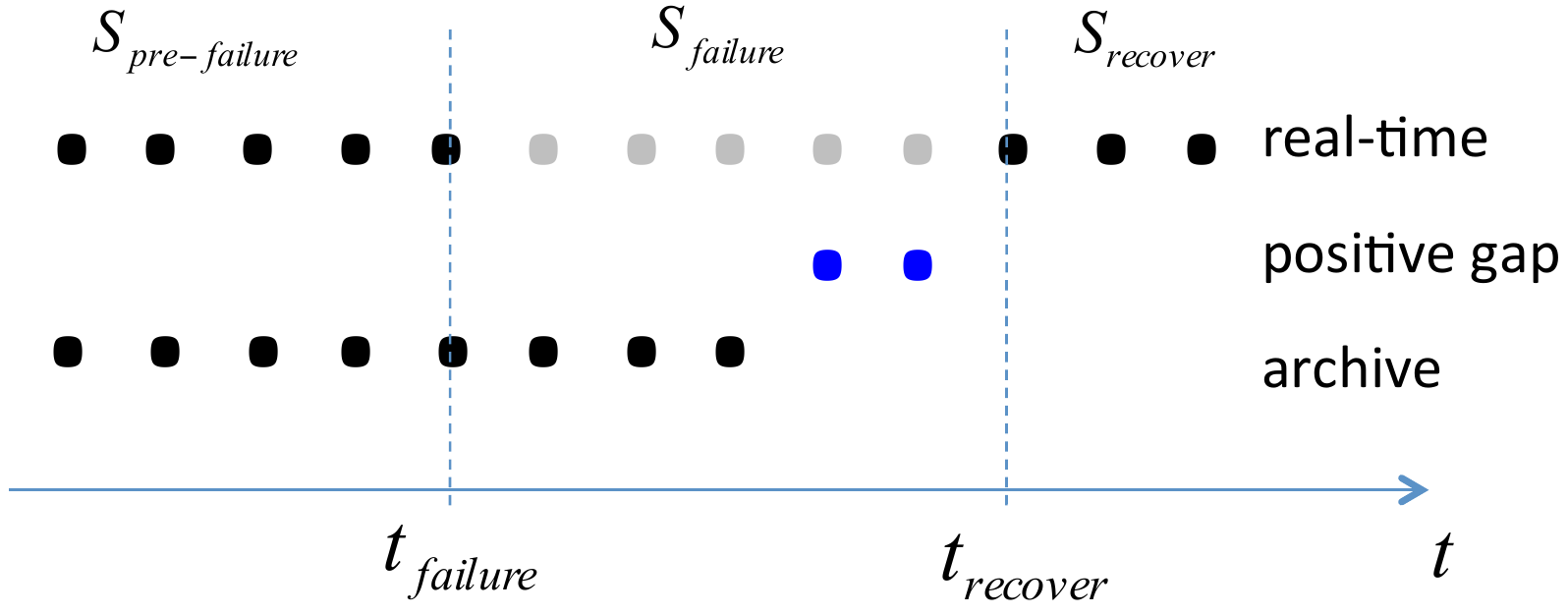}}
\caption{Integrated query plans over end-to-end event streams.}
\label{fig:flowquery}
\end{figure*}

When streams have a \emph{negative gap} between real-time and archived events (Fig.~\ref{fig:negGap}), some events observed by the real-time query engine after query initiation time $t_0^{S}$ are already available in the archive at time $t_0^{S}$. This can lead to duplicate matching patterns to be present in $R = R_{S} \cup R_{D}$. We handle this by adding a filter clause to the archive query so that it does not consider events with timestamp greater than $t_0^{S}$, which are shown as gray points in Fig.~\ref{fig:negGapquery}, but otherwise use the same query plan as \emph{zero gap}.

With a \emph{positive gap} between events in the real-time and the persisted streams (Fig.~\ref{fig:posGap}), we have an ``in-flight'' event-set $M =\left\{e_i \mid i=(-j+1) \ldots -1\right\}$ with events that were already seen by the real-time query engine but not yet in the event database at time $t_0^{S}$. When applying query $Q$ at $t_0^{S}$, these events are neither visible to the real-time query engine nor to the database, causing patterns to be missed. We avoid this situation as follows. For queries without a \texttt{WINDOW} clause, at time $t_0^{S}$ we execute $Q$ on stream $S$ to detect the pattern set $R_{S}$. We wait for a time period $t_0^{S} - t_0^{D}$ and lazily execute $Q$ on the database to get a result set $R'_{D}$. The integrated result set is given by $R = R_{S} \cup R'_{D}$. For queries with \texttt{WINDOW} clauses, the query plan is modified similarly to the zero gap case to consider missed patterns in windows that span the boundary between real-time and archived events. 

\subsection{Query Plan for Fault Resiliency}
\label{sec:ha}
One of the objectives of \kcep queries over past and present events is to adapt to failure conditions. Event processing systems can face hardware or software failures, which, if unaddressed can lead to inaccurate pattern detection. Such systems are often used in mission-critical, low-latency applications that require high availability and consistent results. This is achieved both by mitigating failures and by providing \emph{fail-fast} recovery mechanism. While significant work has gone into making databases persistent and available, say using replication and hot-standby, in-memory CEP systems are prone to faults causes by hardware failures when an engine may go down and a real-time stream may not be visible to it briefly. 

Assuming that the archive database is persistent and reliable, our goal is to extend these resiliency guarantees to the real-time processing engine too. For this, we use the fact that all events are reliably persisted to the archive to provide fault-tolerance for the combined system spanning real-time and persistent streams. The approach we use is as follows. Faults in the (physical or virtual) machine running the real-time engine are detected rapidly in a fail-fast manner. Once detected, the same or a different machine is brought online in O(minutes), with the only state required on the restored machine being the queries registered with the real-time engine and the timestamp of the last event processed before the failure. Meanwhile, all events are reliably being archived to the database without any loss. All of these are reasonable assumptions that can be achieved using various means. The goal for us is to recover the missing patterns that should have been matched by the registered queries during the system downtime and resume processing future events, thus returning a \emph{consistent} (albeit delayed) result stream to the user. 

Fig.~\ref{fig:ha} shows the real-time event stream on top with events being missed by it starting from time $t_{failure}$ until $t_{recovery}$, while at the same time, all events are being durably archived to the database in the bottom row. The failure recovery problem can be reduced to the end-to-end event stream query problem. Here, we consider two boundaries between the real-time and archived events: $t_{failure}$, the time at which the real-time engine fails, and $t_{recover}$, the time when the real-time engine is back online. There are three time segments in which events can exist in the real-time stream: event set $S_{failure}$ which is not observed by the real-time query engine during its downtime, $S_{pre-failure}$ which is processed by the real-time engine before its failure, and $S_{recover}$ which is processed by the real-time engine after it resumes. Assuming a resilient database, $S_{failure}$ is still accessible from the event archive, though there may exist positive, negative or zero gap at the boundary time points $t_{failure}$ and $t_{recover}$ (though a positive gap is the most likely). 

Given a query $Q$ submitted to at $t_0^{S}$, the objective of failure recovery is to reconstruct a result set $R$ which is the same as the expected query result of $Q$ if the system had not failed.  Let results detected by the real-time query engine on $S_{pre-failure}$ be $R_{pre-failure}$, patterns detected on $S_{recover}$ be $R_{recover}$, and patterns retrieved from the archive $S_{failure}$ be $R_{failure}$. In the simple case, the query $Q$ does not have a \texttt{WINDOW} clause and there is zero gap between the real-time and archive data at $t_{failure}$ and $t_{recover}$. The reconstructed query result is:
\[ R=R_{pre-failure}\cup R_{recover}\cup R_{failure} \] 
If $Q$ has a \texttt{WINDOW} clause, let $R_{boundary}^{failure}$ and $R_{boundary}^{recover}$ be the missing patterns at the boundary of $t_{failure}$ and $t_{recover}$, respectively. The complete pattern set is then:
\[ R=R_{pre-failure}\cup R_{recover}\cup R_{failure} \cup R_{boundary}^{failure} \cup R_{boundary}^{recover}\] 
where $R_{boundary}^{failure}$ and $R_{boundary}^{recover}$ can be computed using approaches discussed in the previous sections.

\section{{SCEPter} System Architecture}
\label{sec:arch}

\begin{figure*}[tb]
  \centering
  \includegraphics[width=0.9\textwidth]{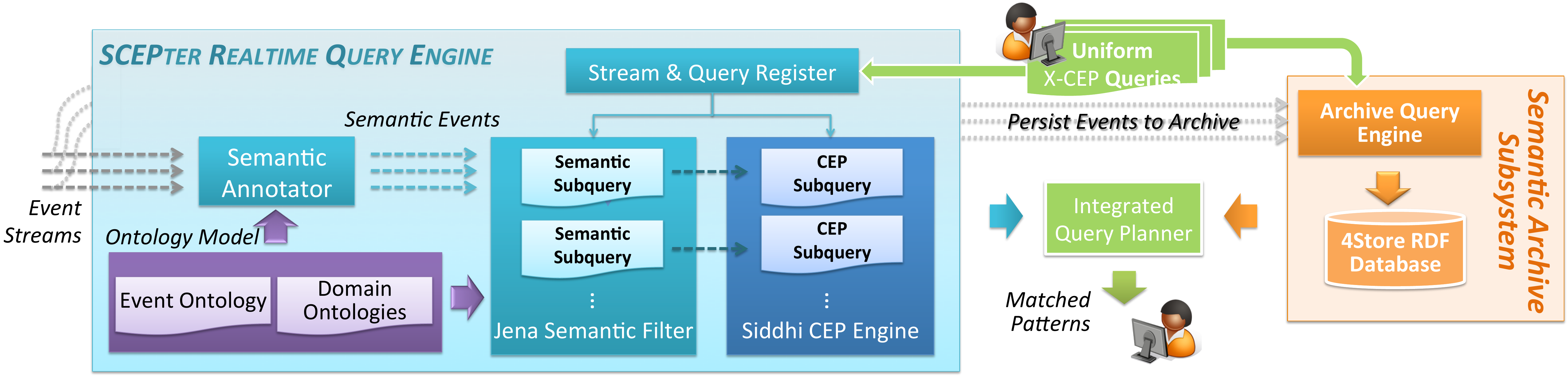}
  \caption{Architecture of the \scep prototype system that implements the \kcep event and query models. Raw events arriving on streams are semantically enriched, pipelined through a semantic filter and a CEP engine for an asynchronous execution using the hybrid strategy, and archived onto an RDF database to support \kcep queries over the end-to-end stream.}
  \label{fig:arch}
\end{figure*}

\scep is our implementation of a semantic complex event processing system over end-to-end event streams that incorporates the \kcep event and query models we have proposed, and validates the query planning and optimizations we have discussed. \scep is implemented fully in Java and consists of three parts -- the real-time query engine, the semantic archive subsystem, and the query planner, as shown in Fig.~\ref{fig:arch}.

\textbf{Real-time Query Engine:} The real-time query engine is built around an existing open-source CEP engine, \emph{WSO2 Siddhi}~\cite{website:Siddhi, jpdc:wihidum}. Siddhi is an exemplar of a traditional CEP engine and uses a tuple-based event model to support common CEP operators for performing filtering, sequence and aggregation patterns over event streams. \scep complements Siddhi to support knowledge-infused semantic event query processing using several modules: \emph{domain ontology model}, \emph{stream and query registry}, \emph{semantic annotator}, and \emph{semantic filter} (Fig.~\ref{fig:arch}, left, in light blue). Event streams are pipelined through the annotator, semantic filtering and Siddhi engine to operate asynchronously.

The \emph{domain ontology model}, represented in OWL, forms a knowledge-base of concepts and their relations to support semantic modeling and query processing. For the microgrid domain, we organize ontologies in a modular fashion for extensibility, as described in our prior work~\cite{zhou2012ITNG}. The \emph{semantic annotator} translates raw event tuples, arriving at runtime, to semantically enriched events based on an annotation file that describes direct mappings from event tuple attributes to semantic properties. The \emph{stream and query register} module processes static event semantics for event schemas and \kcep queries provided by users as discussed in \S~\ref{sec:compile}. 
The \emph{semantic filter} processes dynamic event semantics. Domain ontologies are loaded once into a Jena~\footnote{Jena Framework, \url{jena.sourceforge.net}} in-memory RDF database and is used to evaluate semantic subqueries of registered \kcep queries over incoming events, by a Java thread. If an event satisfies the query, it will be pipelined to the input stream that feeds into the Siddhi CEP engine; otherwise, the event is dropped. The semantic filter incorporates our event buffering and cache optimizations for improved semantic event processing. Finally, the \emph{Siddhi CEP engine} processes CEP subqueries present in the \kcep query. The interfaces to the engine are modular to allow other CEP engines to be used instead, if they provide better capabilities.

\textbf{Semantic Archive Subsystem:} The archive subsystem is used to persist a fork of the incoming event streams to a semantic database and manage hybrid \kcep queries over them (Fig.~\ref{fig:arch}, right, in orange). We use the \emph{4Store RDF database}~\cite{Steve2009} as our persistent backend due to its relative storage scalability and insert performance~\footnote{Even though semantic databases are known to have limited scalability (which is one of the motivations for our hybrid query execution using the CEP engine too), 4Store was among the best performing, non-proprietary RDF stores that was available at the time of writing. As with the CEP engine, 4Store can be replaced by any other RDF database supporting SPARQL.}. 4Store offers a SPARQL REST service for event insertion and querying. The \emph{archive query engine} creates SPARQL queries from registered \kcep queries, and implements the hybrid query evaluation using both query rewriting and execution over the 4Store archive, and the replay of its results for further CEP processing. 

\textbf{Integrated Query Planner:} The query planner coordinates between the real-time query engine and the archive subsystem to retrieve consistent and integrated query results based on the query range specifications and stream configurations (Fig.~\ref{fig:arch}, center, in green). When a user registers a \kcep query, the archive query is generated, and executed immediately or lazily depending on the gap width between the real-time and archived events. The planner then buffers and combines the independent results from the engines into a consistent, ordered stream of query matches for the user to consume.

\section{Performance Evaluation}
\label{sec:experiments} 
The proposed \kcep query model has been implemented within \scep. We have practically verified its feasibility for representing the diverse ontologies required by the Smart Grid domain and defining the various DR query patterns over semantically enriched events, using the real-world USC microgrid IoT deployment. The \kcep model offers a flexible and intuitive query language for specifying powerful knowledge-infused query patterns over this multi-disciplinary domain.  At the same time, as we have noted, the practical use of the query model depends on its ability to be implemented and scaled to high event input rates as observed in IoT domains, and this was the motivation for the various processing optimizations we have proposed. Now, we empirically validate the performance benefits of these optimization techniques that are incorporated in \scep. 

These experiments use the same set of example queries introduced earlier, with ontologies from our prior work~\cite{zhou2012ITNG}, and apply these to real event streams from the USC campus microgrid. Raw event data is collected from 20 airflow sensors in the campus HVAC system to create a corpus of $\sim$120K events, that are then used to simulate input event streams for \scep. This protects the operational microgrid applications during the experiments while providing a realistic scenario, with the added ability to reproducibly evaluate different event rates. The gap width between archive and real-time engines is currently set statically, but is configurable. 

We run \scep on a 12-core 2.8~GHz AMD Opteron server with 32~GB of physical memory, running Windows Server 2008 R2 and using 64-bit Java JDK v1.6. The 4Store database runs within a Linux Virtual Machine (due to OS dependency) on the same machine with exclusive access to 1~CPU Core and 8~GB of RAM, and is accessed by \scep over the local network port. All experiments were performed three times and the average values are reported here.

\subsection{Real-time Query Processing}
In the first set of experiments, we evaluate the time performance of \scep's real-time query processing engine using \kcep queries from \S~\ref{sec:eventpattern}. Specifically, we study the \emph{throughput} of the system and the \emph{processing latency} per event as the input event rate increases, for the following combinations of optimizations and relevant queries that benefit from them: 
\begin{enumerate}
\item Queries with only CEP subqueries, i.e., Queries~\ref{q:simplecep}--\ref{q:seqcep}. This offers a benchmark of the upper bound on event rates that can be processed just by syntactic CEP querying. 
\item Full \kcep queries having both semantic and CEP subqueries, i.e., Queries~\ref{q:simplescep2}--\ref{q:seqscep}, and without any optimizations. This gives a baseline on the lower bound of event rates that can na\"{i}vely be processed using both semantic and syntactic query processing.
\item Full \kcep queries, with only event buffering optimization. We use two different buffering window time durations of $1$~second and $2$~seconds.
\item Full \kcep queries, with only query caching optimization. We use two different cache capacities of $5\%$ and $25\%$, which indicate the fraction of unique cache keys on the input event stream that are retained in the cache.
\item Full \kcep queries with both buffering and cache optimizations enabled.
\end{enumerate}
We observe that the real-time performance is similar for queries with the same semantic subqueries but different CEP subqueries due to the asynchronous pipeline processing architecture and the semantic query processing dominating. So, for brevity, we report only results for Queries~\ref{q:simplecep} and \ref{q:simplescep2}, that are representative of syntactic CEP and \kcep query behavior, respectively.

\begin{figure}[t]
\centering
  \subfloat[Real-time Query Processing Throughput]{\label{fig:throughput}\includegraphics[width=0.47\textwidth]{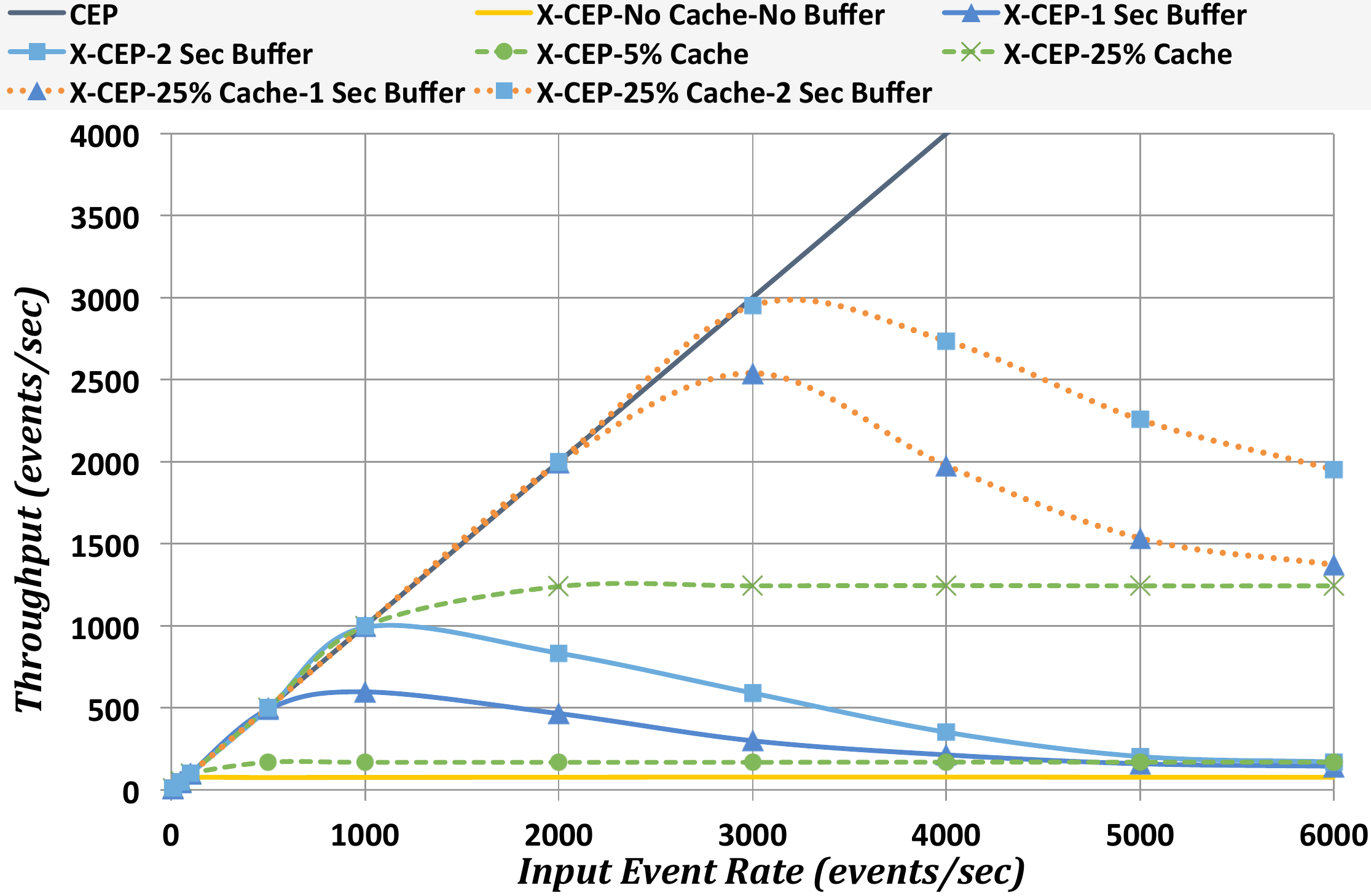}}~~~~
  \subfloat[Real-time query processing latency by module, for different configurations. Semantic annotator and CEP kernel times approaches $0$ms.
]{\label{fig:latency}\hspace{0.1in}\includegraphics[width=0.49\textwidth]{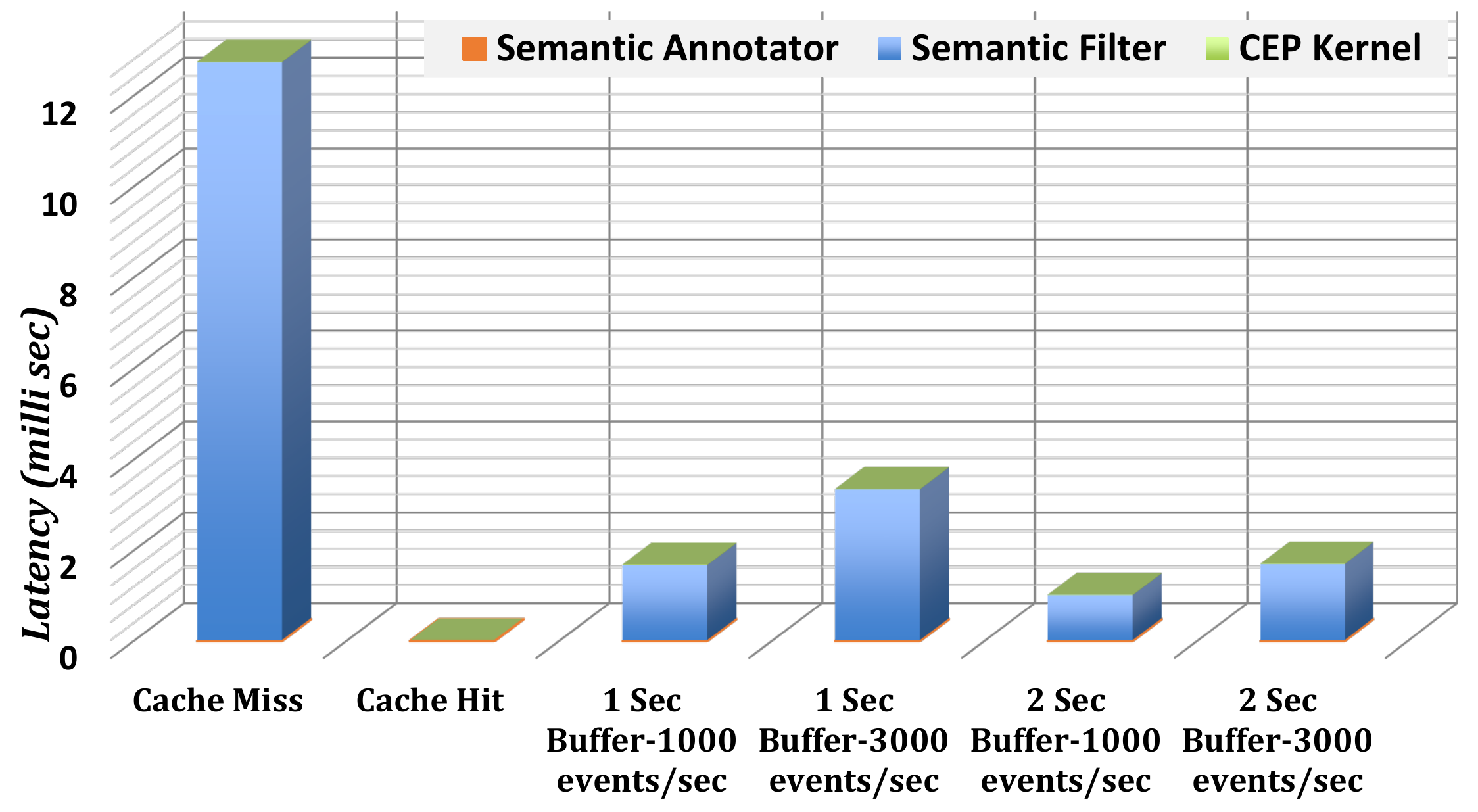}}
\caption{Throughput and latency of CEP and \kcep queries on \emph{real-time streams}, with various combinations of optimizations.}
\end{figure}
Fig.~\ref{fig:throughput} shows the real-time query processing throughput (events/sec) on the Y Axis and the input event rates on the X Axis for CEP Query~\ref{q:simplecep} and \kcep Query~\ref{q:simplescep2}. Ideally, the output throughput should match the input event rate, which occurs only for the \emph{CEP query} (solid black line) where the CEP engine can keep up with input event rates till $100,000$~events/sec that we tested.  For full \kcep queries, the \emph{baseline approach} has a low peak throughput of just $\sim80$~events/sec (solid yellow). This highlights the vast difference between the CEP benchmark and the \kcep baseline, caused by the high semantic processing overhead.

With \emph{event buffering} optimization, the performance is dependent on the input event rate, as we posited earlier -- increasing the buffer window size from 1~sec (solid blue triangle) to 2~sec (solid blue square) improves the peak sustained throughput from $\sim500$~events/sec to $\sim1,000$~events/sec. But it is unable to grow past this peak when the input rate increases further since the time to batch-process the buffer contents eventually outstrips the time to fill the next batch for processing. 

As a note, once the processing throughput of the engine drops below the input rate, any subsequent increase in input rate causes the processing throughput to plateau out or, typically, drop since the system is overloaded with just receiving the additional events at a higher rate. This inflection point when the processing rate diverges from the input rate is the \emph{peak sustained processing throughput}.

With just \emph{caching} enabled, the performance improvement over the baseline is more modest. A cache capacity of $5\%$ (dashed green circle) shows negligible improvement in increasing the processing rate to $170$~events/sec while a larger $25\%$ capacity (dashed green cross) shows a tangible improvement to $\sim1,200$~events/sec. We do see that with the caching optimization, the throughput benefit is retained as the input rate increases, rather than drop down like for buffering.

However, it is with a blend of these two optimizations that we are able to observe a much better performance over the baseline.  Combining \emph{buffering and caching} compounds their performance benefits (dotted orange lines) and we get a sustained throughput of $3,000$~events/sec with a $2$~sec buffer and a $25\%$ cache capacity. Given that the USC campus IoT deployment -- comparable to a small-scale ``smart township'' -- has about $50,000$ infrastructure sensors across $170$ buildings that emit events with a period of between $1$--$15$~mins, this translates to a peak input event rate of about $830$~events/sec, and this can be comfortably managed with our optimizations enabled. If we consider just the smart meters event streams from the entire city of Los Angeles, with about $4M$~consumers emitting data every $15$~mins, this processing capacity we can support is close to the sustained input event rate of $4,400$~events/sec that would be required.

We drill down into the \emph{latency time} spent within each module in the \scep pipeline. Fig.~\ref{fig:latency} shows the average latency of the semantic annotator, semantic filter and the CEP engine for different optimizations. The first two columns show the latencies when only the cache optimization is enabled. When the query cache is missed, we observe that the time is predominantly spent in the semantic filter module, taking about $12$~ms per event -- which translates to about $80$~events/sec in throughput -- while the relative times for annotator and CEP engine are negligible. This confirms that the semantic queries are the most expensive operation. When there is a cache hit, all three times are negligible. {Hence in Fig.~\ref{fig:throughput} the query throughput increases as the cache capacity increases (dashed green circle versus dashed green cross) since the average event processing latency decreases due to more cache hits for the semantic subquery evaluation.}

The last four columns show the latencies when only the buffer optimization is enabled, and caching disabled. When the input event rate increases, the number of events accumulated in a buffer window increases, causing the batch processing time and the per-event latency to increase. If the buffer processing time is within the event rate, the system can keep up with the input stream. For example, with a 2~secs buffer and $1,000$~events/sec input rate (column 5 in Fig.~\ref{fig:latency}), the per-event latency is less than 1~ms and it takes within 2~secs to process the $2,000$~events in the entire 2~sec buffer. But with a 2~sec buffer and $3,000$~events/sec rate, the per-event latency is 1.7~ms and it takes 10.2~secs to process the $6,000$~events that accumulate in the 2~sec buffer window. As a result, the real-time processing cannot keep up with the input rate.

\subsection{Integrated Query Processing}
\begin{figure*}[t]
\centering
  \subfloat[Initial Recovery Latency]{\label{fig:eval3}\includegraphics[width=0.33\textwidth]{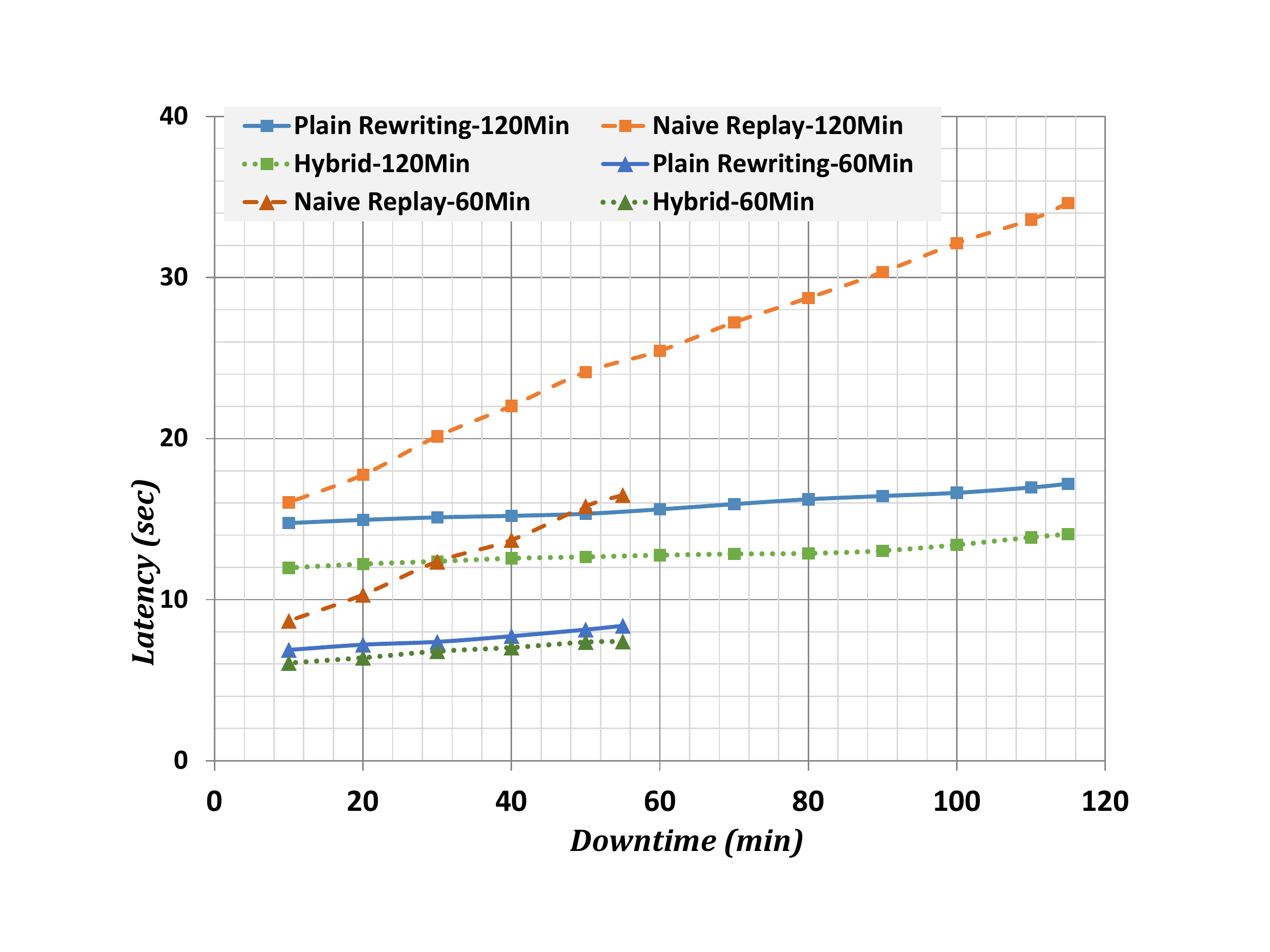}}~
  \subfloat[Catchup Duration]{\label{fig:eval4}\includegraphics[width=0.33\textwidth]{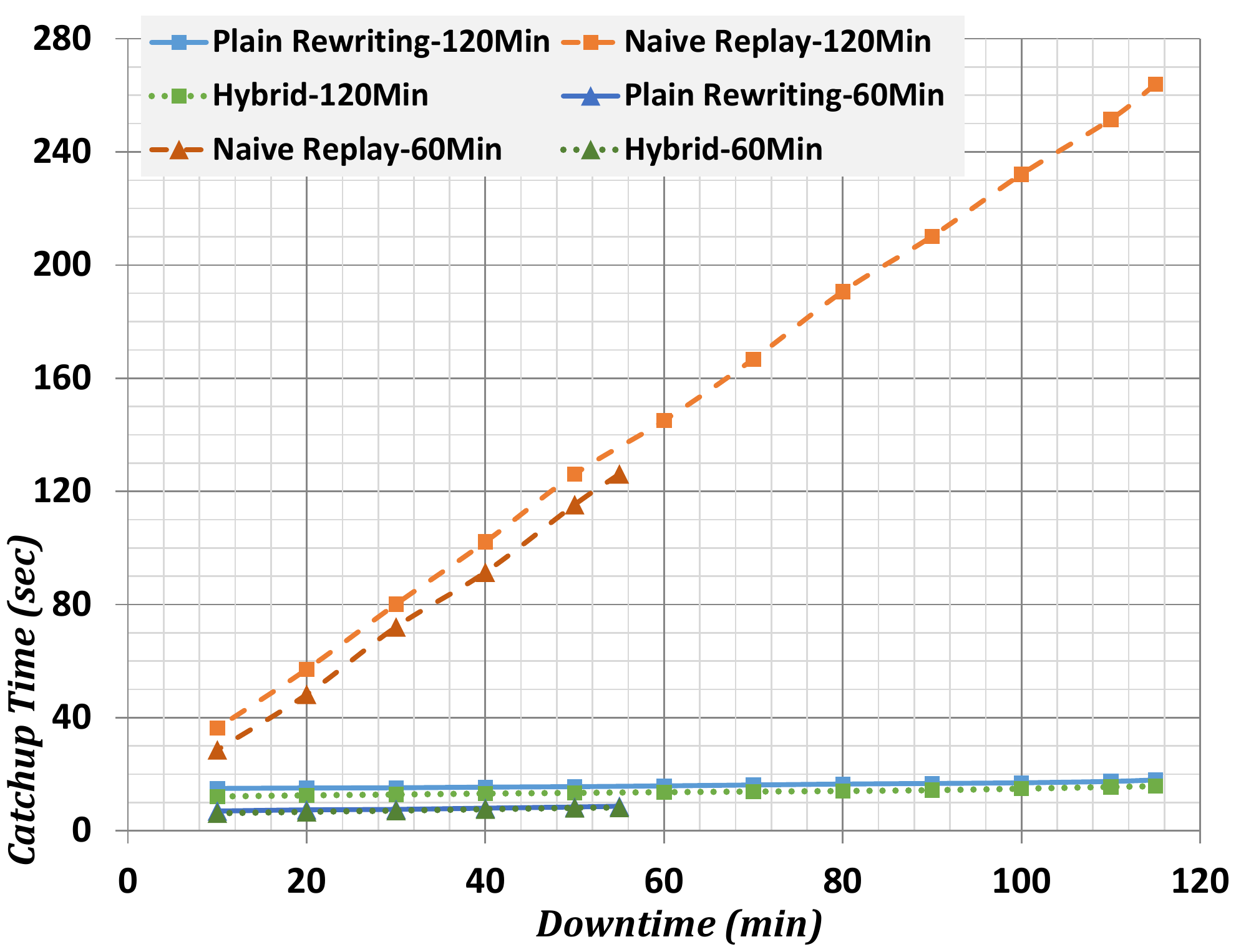}}~
  \subfloat[Catchup Throughput]{\label{fig:eval5}\includegraphics[width=0.33\textwidth]{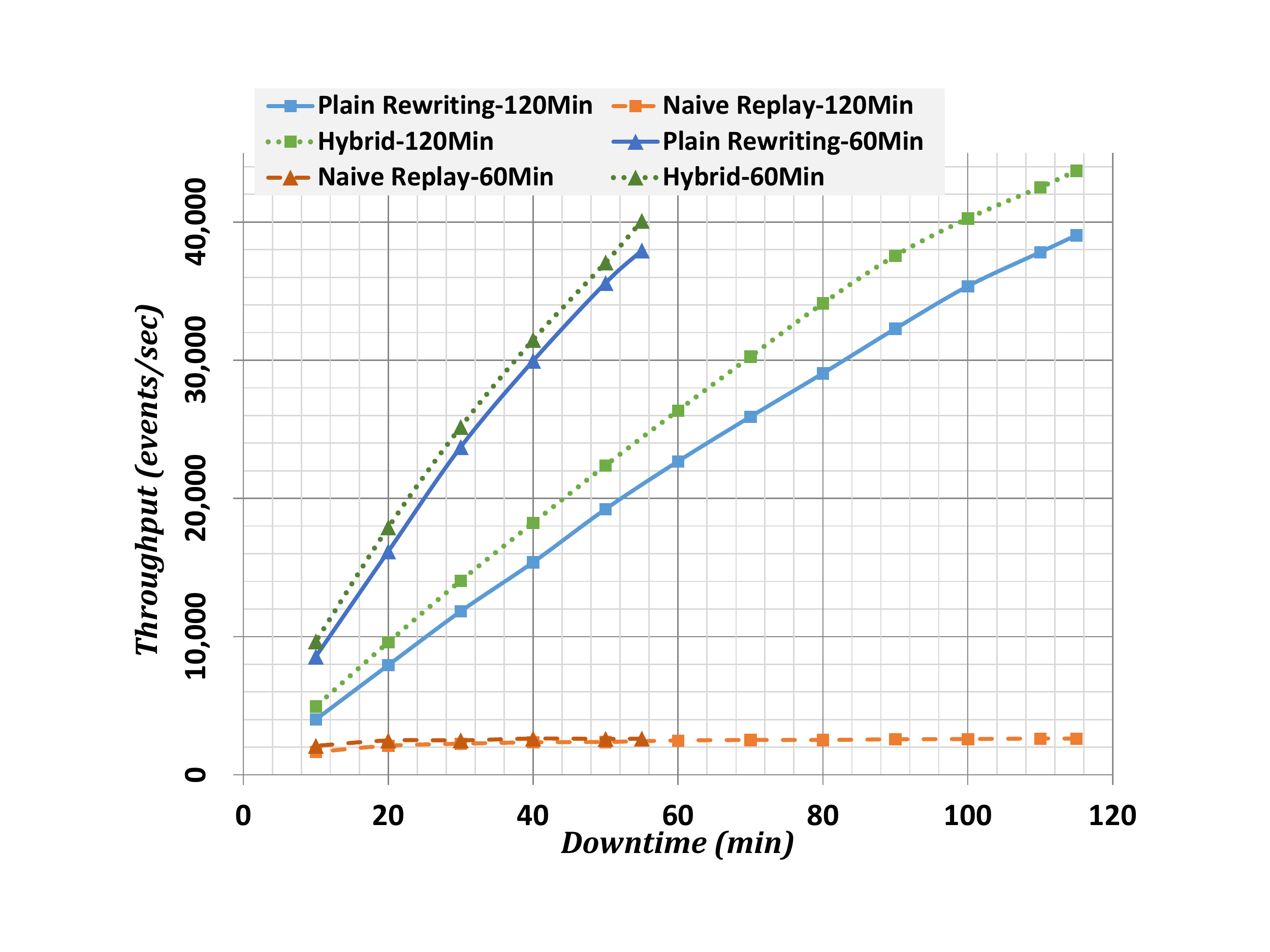}}
\caption{\scep recovery from input stream downtime using query processing on real-time and persistent streams.}
\end{figure*}
In these experiments, we evaluate \scep's performance when performing integrated queries across real-time and persisted event streams. Specifically, to highlight the system performance in the presence of temporal gaps, we introduce a ``downtime'' in the real-time input event stream to \scep, from which it recovers using integrated querying on the event archive and real-time stream. We measure three different metrics: the \emph{initial recovery latency}, the time to process the first archived event after the fault; the \emph{catchup duration}, the time to process all events archived during the downtime; and the \emph{catchup throughput}, the rate of processing the archived events. Once the recovery is complete and the system has caught-up, the real-time stream processing resumes and the results from the previous sub-section apply. 

We compare these metrics for the different archive processing strategies: event replay, plain query rewriting and the hybrid approach, and for stream downtimes that range between 10~mins--115~mins. We also examine the impact of the \emph{4Store database}'s archive size on the performance by using two different archive capacities, one that stores the last 60~mins of input events and the other the last 120~mins of events. Note that the rolling archive capacity determines both the duration for which the real-time engine can fail and recover its result stream consistently, without event loss, and also limits the duration back-in-time for which users can specify their archive query. 

Figs.~\ref{fig:eval3}, \ref{fig:eval4} and \ref{fig:eval5} plot these three recovery metrics for Query~\ref{q:simplescep2}, with HVAC sensor events generated at a rate of $600$~events/sec on the input stream. We see that a \emph{na\"{i}ve replay} approach takes the longest time to recover (orange dashed lines in Fig.~\ref{fig:eval3}), taking 15~secs for the initial recovery from a 10~min downtime, using a 120~min archive, and 36~secs to catch up (Fig.~\ref{fig:eval4}). We also see the impact of larger archive sizes, whereby the smaller 60~min archive takes only 9~secs to recover. This reflects the limited indexing ability of the semantic database, even to perform a temporal range lookup, thus causing costly table scans. Alternatively, if using a pure replay strategy, it will be more efficient to use a relational database with an index on the timestamp column. Note that it is not possible for a 60~min archive to recover from downtimes greater than 60~mins, and hence they are not shown in the figures.

Both plain \emph{rewriting} (blue solid lines) and \emph{hybrid} (green dotted lines) approaches perform all/most of their \kcep queries in the database. So their initial recovery times almost equal their catchup durations -- the real-time query engine makes little/no effort. The hybrid approach has a marginally smaller catchup duration by leveraging the native benefits of both the CEP and the semantic database query engines. As the downtime increases, the na\"{i}ve replay gets progressively worse since it extracts even larger numbers of events from the database for processing by the real-time \kcep engine. This impact is much smaller in the hybrid and plain rewriting approaches, as seen by their flatter slopes in the Figs.~\ref{fig:eval3} and \ref{fig:eval4}, allowing them to scale better.

While not shown in the plots, both the na\"{i}ve replay and the hybrid approaches perform well for queries that also have CEP subqueries since the events are materialized from the database and processed by the CEP query engine which, as we have seen, scales well to large event rates. However, plain rewriting transforms CEP subqueries to SPARQL that is fully executed in the semantic database. We report that for \kcep queries with \texttt{WINDOW} clauses, like Queries~\ref{q:aggscep} and \ref{q:seqscep}, the plain rewriting approach performs much worse, with recovery latencies greater than 10~mins for even short downtimes. As discussed in \S~\ref{sec:dbprocessing}, while it is possible to rewrite correlation constraints to SPARQL, current semantic database implementations map these to repeated and costly self-joins over the entire data set. Increasing the archive capacity exacerbates this. The hybrid approach side-steps this issue by pushing CEP subqueries to the real-time query engine.

\section{Related Work}
\label{sec:related}
Our work falls in the space of stream and Complex Event Processing~\cite{Abadi2005, Demers2007, Akdere2008}. These systems offer intuitive languages to query over continuous data streams with low latency, such as the \emph{boxes-and-arrows} workflow paradigm of Borealis~\cite{Abadi2003}. CEP systems~\cite{Demers2007, Akdere2008}, as an evolution from relational databases, typically follow a SQL-like query syntax and allow explicit representation of temporal patterns, such as sequences, in addition to relational patterns. Existing CEP systems have focused on optimization of temporal pattern matching. Cayuga~\cite{Demers2007} leverages an eager nondeterministic infinite automata (NFA) algorithm to match event sequences incrementally. T-REX~\cite{Cugola2012} compares eager and lazy (buffering) evaluation of CEP queries on real-time streams. \cite{Rabinovich2011} discusses query rewriting techniques to transform common CEP patterns into subpatterns for parallel execution, where possible. \scep is a natural extension of traditional CEP systems where CEP engines are used to process CEP subqueries, and we further support data variety, through semantics, and volume through querying persistent streams. 

One approach to addressing data variety in streams is the utilization of schema mappings~\cite{Mozafari2013}. However, defining one-to-one structural mappings is unsustainable for multi-disciplinary domains with fast changing information spaces, like emerging IoT application in general and Smart Grids in particular. Others have used semantics to manage diverse data in CEP~\cite{Barbieri2010, Anicic2012, Teymourian2010}. Specifically, C-SPARQL~\cite{Barbieri2010} extends the SPARQL language with window and aggregation clauses to support RDF stream processing. However, while it allows time based RDF data filtering and aggregation, it misses a few other basic CEP operators and patterns such as negation and length window. ETALIS~\cite{Anicic2012} is a rule-based deductive system that acts as a unified execution engine for temporal pattern matching and semantic reasoning. It implements two \emph{separate} languages for pattern detection and semantic reasoning: ETALIS Language for Events (ELE), and EP-SPARQL for stream reasoning. Both are transformed to Prolog rules and executed by a Prolog inference engine. Rather than adopting a bespoke solution that departs from traditional CEP systems, our \kcep query model integrates the native and well-understood CEP with semantic knowledge constructs. In practice, this allows our framework to process semantic-enriched CEP queries using existing CEP tools that scale and the proposed optimizations improving performance for semantic processing. 

It is appealing to achieve integrated querying over real-time and persistent streams using a single database engine. Real-time databases iteratively process transactions over constantly changing data, imposing limits on their processing latency. Techniques like scheduling, buffer and cache management~\cite{Kao94, kang:tc:2012} are used to manage temporal consistency and deadlines for returning results. Active databases are another extension to process time-varying data. ECA rules and trigger mechanisms~\cite{Widom1990} were defined to support standing queries along with optimizations for continuous queries. Tapestry~\cite{Terry1992} converts a standing query in active database into an incremental query that finds new matches to the original query as data is added to the database. However, such relational database engines that span persistent and real-time tuples using schedules and triggers are challenged even by the more expressive syntactic CEP queries that can specify temporal constraints, and especially perform poorly for windowed correlations. Semantic queries that are much more complex than CEP queries are infeasible on these platforms. 

Using time windows to correlate and process continuous queries across real-time and archived data has been recently proposed. DataCell~\cite{Liarou2009} exploits relational models specifically for stream processing. Incoming data tuples are cached into ``baskets''  (in-memory tables), queried in batch and flushed from these temporary tables to the underlying database. The basket concept resembles the window operator in CEP queries. This approach potentially allows unified querying on real-time streams and persistent data, but is distinct from our \kcep model where we treat the database as a logical extension of the stream, back-in-time, rather than a static data source to perform a join. Pattern correlation queries (PCQ)~\cite{Dindar2011} define the semantics of a recency-based CEP model over live and archived event streams. The recency clause in PCQ is essentially a \emph{happened-before} relation which specifies the temporal distance between patterns in the live streams and in the persistent streams. It focuses on correlation of patterns which individually consist of either real-time or persisted events. Our system, however, considers a superset of the problem, proposing uniform queries processing approaches across streams with even a temporal gap at the boundary between real-time and persisted streams.

There has been significant work on scalable Big Data platforms for distributed stream processing that complements the work on CEP engines. Systems like Apache Storm~\footnote{Apache Storm distributed real-time computation system, http://storm.apache.org/} from Twitter and Apache Spark Streaming~\cite{zaharia2012discretized} offer the ability to compose a set of user logic blocks that can operate in input streams of events. While Storm offers an imperative model to compose topologies as Directed Acyclic Graphs (DAGs), Spark Streaming also allows limited declarative capabilities using some built-in operators such as aggregation. However, a key distinction between these stream processing systems and complex event processing systems is that the former is not cognizant of the schema of the events and consequently does not offer a declarative query model on which to operate over the input events. The user provided logic blocks are opaque to the DSP platforms. CEP engines, on the other hand, raise the abstraction for users to specify their event pattern, including domain knowledge, in the case of \kcep, and transparently execute them. In fact, there has been work such as Apache Kafka and even in Siddhi to leverage DSP platforms to support scalable CEP queries~\cite{storm-cep}.

\section{Conclusions}
\label{sec:conclusions}

In this article, we have introduced a novel \kcep query model to enhance CEP queries with knowledge semantics, and also perform such queries seamlessly over end-to-end event streams, from network to disk. This unified real-time analytics model allows users to easily specify event patterns over diverse knowledge concepts in emerging domains, such as IoT and Smart Cities, and on streaming information that spans past, present and future. We have illustrated the value of this model using a case study from the Smart Grid IoT domain, and highlighted its ability to address gaps in the current state-of-the-art to support the critical needs of real-time knowledge-infused analytics for multi-disciplinary domains. As a result, we address the velocity, variety and volume dimensions common to Big Data applications.

Further, we have proposed approaches to translate costly semantic subqueries in the \kcep model into scalable execution, with optimizations to address the current limitations of semantic engines to process queries with low latency and on large archives. These have been implemented in the \scep system prototype to validate our design, and to leverage the best capabilities of existing native CEP and RDF database engines. Our experiments confirm that our cache and buffer optimizations are able to achieve a \kcep query processing throughput of $3,000$~events/sec, which is a $30\times$ improvement over the baseline approach that does not include the optimizations, and is adequate to support the needs of a smart township such as the USC microgrid. 

We also show the value of end-to-end processing in not just in allowing users to specify back-in-time queries but also to enhance the fault tolerance of streaming applications. We use the durable and persistent stream archive to consistently recover from transient failures in real-time engines, and map this back to a scenario where there is a temporal gap in processing an end-to-end stream. We are able to recover from outages that last as long as 2~hrs within a $20$~secs latency, and this allows resilient processing of \kcep queries with a modest delay introduced due to the fault.

As the IoT deployments grow and diverse data streams are integrated from multi-disciplinary sources to support customized applications, the value of the proposed \kcep model and its knowledge-infused queries will grow. Its ability to rapidly and meaningfully utilize a variety of event streams, the flexibility to perform seamless analytics backward and forward in time, and take real-time decisions on the IoT system based on pattern matches will prove of critical importance. The consistency and fault-tolerance that our proposed strategies provide makes our system particularly well-suited for IoT applications that require robust guarantees.

\section*{Acknowledgment}

This material is based upon work supported by the United States Department of Energy under Award
Number DOE-0000192 and the National Science Foundation under Award CCF-1048311. The views and
opinions of authors expressed herein do not necessarily state or reflect those of the US Government
or any agency thereof. This work was also supported by grants to the second author from the Department of Electronics \& Information Technology (DeitY), India, and the Robert Bosch Centre for Cyber Physical Systems (RBCCPS), IISc. 

\bibliographystyle{abbrv}
\bibliography{qunzhiRef}

\end{document}